\documentclass[pra,twocolumn,preprintnumbers,amsmath,amssymb]{revtex4}
\usepackage{graphicx}
\usepackage{epstopdf}
\usepackage{amscd}

\usepackage{dcolumn}
\usepackage{bm}

\newcommand{\mment}[1] {\ensuremath{\mathbf{#1}}} 
\newcommand{\interaction}[1] {\ensuremath{\mathbf{#1}}} 

\newcommand{\set}[1] {\ensuremath{\mathcal{#1}}} 

\newcommand{\map}[1]{\ensuremath{\mathcal{#1}}}

\newcommand{\state} [1] {\ensuremath{\mathbf{#1}}} 

\newcommand{\spc}[1]{\ensuremath{\set{#1}}} 

\newcommand{\numberfield}[1]{\ensuremath{\mathbb{#1}}} 

\newcommand{\vect}[1] {\ensuremath{\vec{#1}}} 
\newcommand{\cvect}[1] {\textsf{#1}} 

\newcommand{\lvect}[1] {\ensuremath{\mathbf{#1}}} 

\newcommand{\cmatrix}[1]{\textsf{#1}} 

\newcommand{\rmatrix}[1]{\ensuremath{#1}} 

\newcommand{\Bpi}{\bm{\pi}} 

\newcommand{\model}[1]{\ensuremath{\mathbf{#1}}} 

\newcommand{\hypersphere}{\ensuremath{\mathit{S}^{2N-1}}} 

\newcommand{\var}{\ensuremath{\chi}} 

\newcommand{\vard}{\ensuremath{\xi}} 

\newcommand{\transpose}{\ensuremath{^\text{\textsf{T}}}} 

\addtocounter{secnumdepth}{-1} 

\begin{document}



\title{An Information-Geometric Reconstruction of Quantum Theory,~I: \\
        The Abstract Quantum Formalism}

\author{Philip Goyal}
    \email{pgoyal@perimeterinstitute.ca}
    \affiliation{Perimeter Institute, \\ Waterloo, Canada}

\begin{abstract}
In this paper and a companion paper, we show how the framework of information geometry, a geometry of discrete probability distributions, can form the basis of a derivation of the quantum formalism.  The derivation rests upon a few elementary features of quantum phenomena, such as the statistical nature of measurements, complementarity, and global gauge invariance.  It is shown that these features can be traced to experimental observations characteristic of quantum phenomena and to general theoretical principles, and thus can reasonably be taken as a starting point of the derivation.  When appropriately formulated within an information geometric framework, these features lead to~(i)~the abstract quantum formalism for finite-dimensional quantum systems, (ii)~the result of Wigner's theorem, and~(iii)~the fundamental correspondence rules of quantum theory, such as the canonical commutation relationships.  The formalism also comes naturally equipped with a metric~(and associated measure) over the space of pure states which is unitarily- and anti-unitarily invariant.  
The derivation suggests that the information geometric framework is directly or indirectly responsible for many of the central structural features of the quantum formalism, such as the importance of square-roots of probability and the occurrence of sinusoidal functions of phases in a pure quantum state.  Global gauge invariance is seen to play a crucial role in the emergence of the formalism in its complex form.

\end{abstract}

\maketitle

\section{Introduction}

The formalism of quantum theory has many mathematical features, such as its use of complex vector spaces, whose physical origin is obscure.   Over the last two decades, there has been growing interest in the elucidation of the origin of these features by attempting to derive the quantum formalism from a set of physical assumptions,  motivated by the belief that such an elucidation would contribute to the development of a clearer physical understanding of quantum theory, and might contribute  to the development of a theory of quantum gravity~\cite{Wheeler89, Rovelli96, Popescu-Rohrlich97, Summhammer94, Zeilinger99, Fuchs02}.   

The interest in reconstructing the quantum formalism in this way has been encouraged by belief in the hypothesis that the concept of information might be the key, hitherto missing, ingredient which, if appropriately conceptualized and formalized, might make such a reconstruction possible.    In recent years, several attempts have been made to systematically explore the reconstruction of the quantum formalism from an informational starting point, for example~\cite{Wootters80, Wootters-statistical-distance, Brukner99, Brukner02a, Brukner02b, Summhammer94, Rovelli96, Grinbaum03,
Grinbaum04, Caticha98b, Caticha99b, Clifton-Bub-Halvorson03}.    Perhaps the earliest such work, due to Wootters~\cite{Wootters80}, is remarkable for showing that one can derive a correct, non-trivial physical prediction~(Malus' law) concerning a quantum experiment from a simple information-theoretic principle that concerns the amount of information obtained when measurements are performed upon a system, thereby providing clear support to the idea that the quantum formalism owes at least part of its mathematical structure to quantitative measures of information.  However, the attempt to extend these principles in the direction of the quantum formalism meets with limited success.  Other attempts to quantify the gain of information and to formulate information-theoretic principles have also been made in~\cite{Summhammer94, Brukner99, Brukner02a, Brukner02b}, but they are similarly restricted in their scope.   More recently, a number of approaches~\cite{Rovelli96, Grinbaum03, Grinbaum04, Caticha98b, Caticha99b, Clifton-Bub-Halvorson03} succeed in deriving a significant portion of the abstract formalism of quantum theory, but at the cost of abstract assumptions~(typically involving the assumption of the complex number field) that play a key role~\footnote{
For example,  (a)~Caticha~\cite{Caticha98b}, in a derivation of Feynman's rules for combining probability amplitudes, assumes that a complex
    number can be associated with experimental set-up, (b)~Grinbaum~\cite{Grinbaum04}, in 
    Axiom VII, makes specific assumptions regarding
    the applicable number fields,  and
    (c)~in~\cite{Clifton-Bub-Halvorson03}, it is assumed that a physical
    theory can be accommodated within a~$C^*$-algebraic framework.},
which significantly detracts from the understanding of the physical
origin of the quantum formalism that they can provide.   The approach presented here suggests that it is possible to derive a large part of the quantum formalism without recourse to abstract assumptions of this nature~%
\footnote{The approach presented here is a development of earlier work described in~\cite{Goyal-QT1, Goyal-QT2}.}.

One way to understand the reason for thinking that information might have a fundamental role to play in our understanding of the physical world is as follows.  Suppose that an experimenter is presented with a 
system prepared in a way unknown to her.  In classical physics, she can, in principle, perform an ideal
measurement upon the system which provides complete information about the
state of the system.  We shall describe this situation as one of \emph{transparency}:~intuitively, there is no insurmountable barrier between the system and the observer, so that, \emph{in principle}, there is no fundamental distinction between the state of the system~(the theoretical description of the reality) on the one hand, and the experimenter's \emph{information} about the state on the other.  On the other hand, when the same situation is described using quantum physics, an ideal measurement performed on the system
provides the experimenter with only \emph{partial} information about
the state of the quantum system.  Hence, transparency no longer holds.  Now, one might imagine that the experimenter could bypass this restriction by performing a large number of measurements on an ensemble of identically-prepared systems.  However, if one assumes that an experimenter can, in principle, only perform a finite number of such measurements in a finite time, then, since the information that can be obtained by a finite number of measurements still only provides partial information, it follows that a fundamental distinction is drawn between the state of the system and the information that the experimenter can possibly have about the state.  

It is then natural to ask how the information that the experimenter gains can be quantified, and it seems reasonable to suppose that the information gain, and the manner in which this information is theoretically represented and manipulated, might be constrained by principles which give rise to non-trivial mathematical structure.  This is the line of thinking we attempt to develop in this paper.

In the quantum description of the above situation, the limitation imposed on the information that the experimenter obtains can be traced directly to the fact that, in the quantum case, the state of the system only determines the probabilities of the results of a measurement performed upon it, a property we will refer to as \emph{statistical determination}.  As a consequence, a finite number of runs of an experiment where a measurement is performed on a system prepared in some unknown state only yields limited information about the state.  For example, if the system is in the state~$\cvect{v} = (\sqrt{p_1} e^{i\phi_1}, \sqrt{p_2}e^{\phi_2}, \dots, \sqrt{p_N}e^{i\phi_N})\transpose$ in the basis of a projective measurement,~$\mment{A}$,  where the~$p_i$ are the measurement probabilities and the~$\phi_i$ are phases, then $n$~runs of an experiment where measurement~$\mment{A}$ is performed yield frequencies~$f_1, f_2, \dots, f_N$, with~$f_i$ being the frequency of result~$i$,  which provide only a finite amount of information about the~$p_i$. 

In order to focus upon statistical determinacy and quantitatively explore its consequences, it is helpful isolate this feature, and to consider the process of information gain in the context of classical probability theory.  
Consider the following simple situation. Alice has two coins,~$A$ and~$B$, characterized by the probability distributions~$\lvect{p} = (p_1, p_2)$ and~$\lvect{p}' = (p_1', p_2')$, respectively. Suppose that she chooses coin~$A$, tosses it~$n$ times, and then sends the data to Bob, without disclosing which coin she chose.  If Bob knows~$\lvect{p}$ and~$\lvect{p}'$, how much information does the data provide him about which coin was tossed? 

Using Bayes' theorem and Stirling's approximation for the case where~$n$ is large, on the assumption that coins~$A$ and~$B$ are \emph{a priori} equally likely to be chosen, one finds that~$P_A/P_B =  \exp\left( n \sum_{i=1}^2 p_i \ln (p_i / p_i') \right)$,
where~$P_A$ is the probability that the tossed coin is~$A$ given the data, and likewise for~$P_B$.
When the probability distributions are close, so that~$\lvect{p}' = \lvect{p} + d\lvect{p}$, the argument of the exponent can be expanded in the~$dp_i$ to give~$P_A/P_B = \exp(2n\, ds^2)$, where~$ds^2 = \frac{1}{4}\sum_i dp_i^2/p_i$.

Now, the information gained by Bob,~$\Delta I$, is defined as~$\Delta I \equiv U(1/2, 1/2) - U(P_A, P_B)$, with~$U$ being an entropy~(uncertainty) function such as the Shannon entropy.   But, since~$P_A + P_B = 1$ and~$P_A/P_B$ is determined by~$ds$, once~$U$ is selected,~$\Delta I$ is determined by~$ds$.  This result immediately generalizes to the case where~$\lvect{p}$ and~$\lvect{p}'$ are $M$-dimensional probability distributions~($M \ge 2$).    Hence, from an informational viewpoint, it is natural to endow the space of discrete probability distributions with a Riemannian metric,~$ds^2 = \frac{1}{4}\sum_i dp_i^2/p_i$.  This metric, known as the \emph{information metric}, is the central component of \emph{information geometry}~\cite{Amari85}.   

In the course of this paper and a companion paper~\cite{Goyal-QT2b}~(hereafter
referred to as Paper~II), we show that, by formalizing some elementary features of quantum phenomena~(particularly complementarity and global gauge invariance) within this information-geometric framework, and supplementing these with a few additional plausible assumptions, it is possible to reconstruct the quantum formalism.
In particular, we derive:
\begin{itemize}
\item[1.] The finite-dimensional abstract quantum formalism, namely~(a)~the von
Neumann postulates for finite-dimensional systems, and~(b)~the tensor
product rule for expressing the state of a composite system in terms
of the states of its subsystems.
\item[2.] The result due to Wigner
that a one-to-one map over the state space of a system which represents a symmetry transformation of the 
system is either unitary or antiunitary~\cite{Wigner-group-theory}.
\item[3.] The principal correspondence rules of quantum theory~\footnote{The
correspondence rules of quantum theory can be categorized as
follows:~(i)~\emph{Operator Rules:}~the rules for writing down
operators representing measurements that, from a classical
viewpoint, are measurements of functions of other observables,
(ii)~\emph{Commutation Relations:}~the commutation relationships for
measurement operators, for example those operators representing
measurements of position, momentum, and components of angular
momentum, (iii)~\emph{Transformation Operators:}~explicit forms of
the operators that represent symmetry transformations~(such as
displacement) of a frame of reference, and
(iv)~\emph{Measurement--Transformation Relations:}~the relations
between measurement operators and the operators representing passive
transformations between physically equivalent reference frames.},
such as the canonical commutation relation~$[\cmatrix{x}, \cmatrix{p}_x]=i\hbar$.
 \end{itemize}
In addition, we obtain the unitarily- and antiunitarily-invariant metric~$ds^2 = |d\cvect{v}|^2$, and associated measure, over the space of pure states.

The present paper is organized as follows.  We begin in Sec.~\ref{sec:intro-overview} by providing a brief overview of the main ideas and the main steps in the derivation. 

In Sec.~\ref{sec:F-idealised-set-up}, we formulate a set of
background assumptions and idealizations, and an experimental framework.   The experimental framework precisely delineates which experimental set-ups can be described by the theoretical model that is to be developed.   By proceeding in this manner, we ensure at the outset that there is no ambiguity as to what set of experimental set-ups we are seeking to model.   
In Sec.~\ref{sec:F-statement-of-postulates}, we present a
set of postulates which determine the theoretical model of a system
in the context of this experimental framework.    

In Sec.~\ref{sec:overview}, it is shown how the majority of these postulates can be regarded as reasonable generalizations of elementary experimental facts that are characteristic of quantum phenomena, or are drawn from classical physics, and so can be regarded as a reasonable starting point for a reconstruction of quantum theory.

In Sec.~\ref{sec:D}, we show how these postulates give rise to the finite-dimensional abstract quantum formalism~(apart from the form of the temporal evolution operator) and Wigner's result, as well to a metric and measure over the space of pure states.  We conclude in Sec.~\ref{sec:discussion} with a discussion of the
results.  

In Paper~II, we
formulate an additional principle~(the \emph{Average-Value
Correspondence Principle}), and use this to obtain the form of the
temporal evolution operator and to obtain the correspondence rules of quantum
theory.

\section{Outline of the Derivation} 

\label{sec:intro-overview}

The outline below is presented in four parts.  First, we express the notion of complementarity, which consists of the idea that when a measurement is performed upon a system in some state, the measurement result only yields information about \emph{half} of the experimentally-accessible degrees of freedom of the state.   Complementarity is expressed by the assertion that, in the case of a measurement that yields $N$~possible \emph{results,} each result coarse-grains over two objectively-realized \emph{outcomes} of the measurement.  The outcomes are assumed to be statistically determined by the state of the system, and are thus characterized by a probability distribution.  The assumed statistical determinism of the outcomes motivates the imposition of the information metric over the space of probability distributions.   Using these two ideas, we obtain a formalism in which states of a system are represented by unit vectors in a $2N$-dimensional real Euclidean space, and physical transformations are represented by orthogonal transformations of these vectors.

Second, the state is expressed in terms of  the probabilities of the measurement results and a set of~$N$ real variables,~$\var_1, \dots, \var_N$.  The assertion that the transformation~$\var_i \rightarrow \var_i + \var_0$ is, for any~$\var_0$, is a global gauge transformation of the~$\var_i$ implies that the resulting formalism can be expressed in complex form, such that states are represented by $N$-dimensional complex vectors, and physical transformations are represented by unitary or antiunitary transformations.

Third, it is asserted  that any measurement describable by the formalism can be simulated in terms of any given measurement flanked by suitable interactions, which yields the Born rule and the Hermitian representation of measurements.

Fourth, we assume that, in the special case of a system of definite energy in stationary state whose observable degrees of freedom are time-independent, the temporal rate of change of the overall phase of the state is proportional to the energy of the system.  In conjunction with the global gauge invariance condition, this assumption yields the tensor product rule.  

Finally, we indicate the further steps that are necessary to derive the explicit form of the temporal evolution operator, and the correspondence rules of quantum theory.

\subsection*{Part 1:~Constructing State Space}

Measurement is idealized as a process that (i)~when performed upon some physical system, yields one of~$N$ possible \emph{results,} with probabilities,~$p_1, \dots, p_N$, that are  determined by the state of the system immediately prior to the measurement, and (ii)~is reproducible, so that, upon immediate repetition of the measurement, the same result is obtained with certainty.

\subsubsection{Formalizing Complementarity}  
\begin{figure}[!h]
            \includegraphics[width=3in]{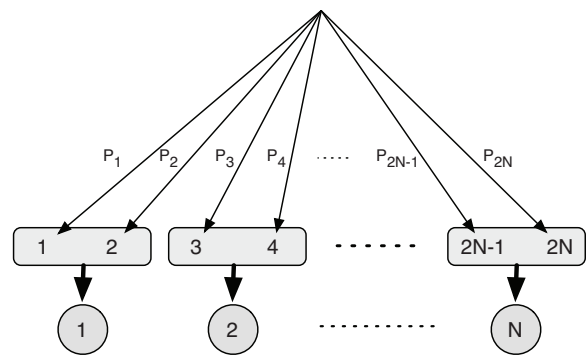}
            \caption{\label{fig:prob-tree-complementarity} \emph{Complementarity Postulate}. Probability
            tree showing the outcomes of measurement~$\mment{A}$, and the corresponding results~(circled) 	
            that are observed.  One of~$2N$ possible outcomes are 	
            realized, with probability~$P_1, P_2, \dots, P_{2N}$, respectively.   The measurement is 
            unable to resolve the individual outcomes.  Result~$i$ is obtained whenever either outcome~$2i-1$ or 
            outcome~$2i$ is realized~($i=1, \dots, N$).}
 \end{figure}

Complementarity is incorporated by assuming that, when measurement~$\mment{A}$ is performed, there are, in fact, $2N$~possible \emph{outcomes}, but these outcomes are not individually observed.  Result~$i$ is observed~($i=1, \dots, N$) whenever either outcome~$2i-1$ or outcome~$2i$ is realized~(see Fig.~\ref{fig:prob-tree-complementarity}).  The probabilities of outcomes~$1, \dots, 2N$ are given by~$P_1, \dots, P_{2N}$, so that~$p_i = P_{2i-1} + P_{2i}$.  The~$P_q$~($q=1, \dots, 2N$) are summarized by the probability $n$-tuple~$\lvect{P} = (P_1, \dots, P_{2N})$.   

Intuitively, performing the measurement brings about the realization of one of $2N$~possible outcomes, but the results of the measurement  \emph{coarse-grain} over these outcomes:~when outcomes~$2i-1$ or~$2i$ is realized, the measurement is~(for some reason to be investigated) unable to resolve the individual outcomes, so that only result~$i$ is registered.

\subsubsection{Imposing the Information Metric}  Once one assumes that measurement outcomes are only statistically determined by the state of the system, it follows that it is impossible in general to determine with certainty whether a given system is in one of two possible states on the basis of the outcomes of a finite number of measurements performed upon identical copies of the system.   Instead, only a finite amount of information about which state is present can be obtained.  Consequently, as discussed in the Introduction, it is natural to endow the space of the~$\lvect{P}$ with the information metric,
\begin{equation}
ds^2 = \frac{1}{4} \sum_q dP_q^2/P_q.
\end{equation}
It is convenient to define~$Q_q = \sqrt{P_q}$, since  the metric over the~$Q_q$ is then simply the Euclidean metric,
\begin{equation}
ds^2  = dQ_1^2 + \dots  + dQ_{2N}^2.
\end{equation}
so that~$\lvect{Q}$ is a unit vector that lies on the positive orthant of the unit hypersphere,~$\hypersphere$, in a $2N$-dimensional real Euclidean space.

\subsubsection{Representing Physical Transformations}  Next, we consider transformations which represent physical transformations of the system.  We seek one-to-one transformations that preserve the information metric over~$\lvect{P}$, thereby endowing the information metric with a fundamental physical significance.  Now, if one takes the~$\lvect{Q}$ themselves as the state space of the system, one finds that non-trivial one-to-one transformations of the state space that preserve the Euclidean metric are not possible.  A simple way to allow the existence of such transformations is to take the entire unit hypersphere,~$S^{2N-1}$, as the state space of the system.   Thus, we shall postulate that a state of the system is given by a unit vector~$\lvect{Q} = (Q_1, Q_2, \dots, Q_{2N})$,  with~$Q_q \in [-1, 1]$, and that the probabilities~$P_q$ are given by~$P_q = Q_q^2$.  From the information metric over the~$\lvect{P}$, it follows that the metric over the~$\lvect{Q}$ is Euclidean.  Hence,~$\lvect{Q}$  lies on the unit hypersphere,~$\hypersphere$, in a $2N$-dimensional real Euclidean space. 

The signs of the~$Q_q$ determine in which of the~$2^{2N}$ possible orthants of~$\hypersphere$ that~$\lvect{Q}$ lies.  We shall refer to the sign of~$Q_q$ as the \emph{polarity} of outcome~$q$, which is defined only when~$P_q \neq 0$, and is physically realized whenever outcome~$q$ is realized, but is unobserved.  

We postulate that any transformation,~$\map{M}$, of~$S^{2N-1}$, that represents a physical transformation of the system is one-to-one and preserves the metric.  It follows that $\map{M}$~is an orthogonal transformation of~$\hypersphere$, so that state~$\lvect{Q}$ is transformed to~$\lvect{Q}' = \rmatrix{M}\lvect{Q}$, where~$\rmatrix{M}$ is a~$2N$-dimensional real orthogonal matrix.

 
\subsection*{Part 2:~Global Gauge Invariance}

We begin by expressing the state,~$\lvect{Q}$,  in terms of the probabilities~$p_1, p_2, \dots, p_N$, and $N$~additional real degrees of freedom,~$\var_1, \var_2, \dots, \var_N$.  In particular, we postulate that the~$p_i$ and~$\var_i$ together determine the~$Q_q$ through the relations~$Q_{2i-1} = \sqrt{p_i} f(\var_i)$ and~$Q_{2i} = \sqrt{p_i} \tilde{f}(\var_i)$, where~$f$ is not a constant function, and~$f$ and~$\tilde{f}$ are differentiable functions to be determined.   

\begin{figure}[!h]
            \includegraphics[width=3.25in]{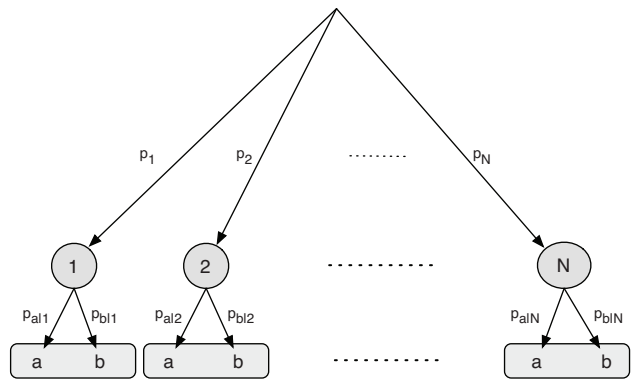}
            \caption{\label{fig:prob-tree} \emph{State Representation Postulate}. Probability
            tree showing the outcomes~(relabeled as~$1a, 1b, \dots, Nb$) of measurement~$\mment{A}$, redrawn to place results~(circled) at the top. Result~$i$ is observed with probability~$p_i$.  Given that~$i$ is observed, outcome~$1a$ or~$1b$ is realized with probability~$p_{a|i} =f^2(\var_i)$ or~$p_{b|i} =\tilde{f}^2(\var_i)$, respectively.   The polarities~(when defined) associated with outcomes~$ia$ and~$ib$ are given by the signs of~$f(\var_i)$ and~$\tilde{f}(\var_i)$, respectively.  The individual outcomes~$ia$ and~$ib$, and their polarities, are unresolved by the measurement.}
 \end{figure}

This change of variables can be understood more directly as follows.  Let the outcomes~$1, 2, \dots, 2N$ be relabeled as~$1a, 1b, 2a, 2b, \dots, Nb$, respectively, where the number~$(1, \dots, N)$ indicates the observed result and the letter~($a$ or~$b$) indicates which of the two possible outcomes compatible with the observed result was realized.  It then follows from the above assumption that, given result~$i$ is observed,  the probability that outcome~$1a$ was realized is~$p_{a|i} = f^2(\var_i)$ and, similarly, the probability that outcome~$1b$ was realized is~$p_{b|i} = \tilde{f}^2(\var_i)$.  The polarities~(when defined) associated with outcomes~$ia$ and~$ib$ are given by the signs of~$f(\var_i)$ and~$\tilde{f}(\var_i)$, respectively~(see Fig.~\ref{fig:prob-tree}).

\subsubsection{Formalizing Global Gauge Invariance}  On the basis of a classical-quantum correspondence argument, we are led to the assumption that, for the model of a system where the state as represented by~$(p_i; \var_i)$, the transformation~$\var_i  \rightarrow \var_i + \var_0$ for~$i=1, \dots, N$ is a gauge transformation, and therefore causes no change in the predictions of the model.   From this assumption, we immediately draw two postulates.  First, that the measure~$\mu(p_i;  \var_i)$, induced by the metric over~$\hypersphere$ is consistent with this global gauge condition, that is~$\mu(p_1, \dots, p_N; \var_1, \dots, \var_N) = \mu(p_1, \dots, p_N; \var_1 + \var_0, \dots, \var_N + \var_0)$ for all~$\var_0$, which we find implies that
\begin{equation}
\begin{aligned}
f(\var_i) &= \cos(a\var_i + b) \\
\tilde{f}(\var_i) &= \sin(a\var_i + b),
\end{aligned}
\end{equation}
where~$a \neq 0$ and~$b$ are constants, so that
\begin{equation}
\lvect{Q} = (\sqrt{p}_1 \cos\phi_1, \sqrt{p}_1 \sin\phi_1, \dots, \sqrt{p}_N \sin\phi_N),
\end{equation}
where~$\phi_i \equiv a\var_i +b$.    

\begin{figure}[!h]
            \includegraphics[width=2.5in]{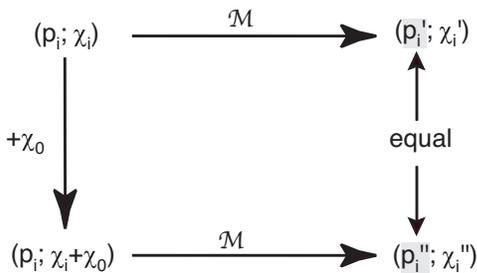}
            \caption{\label{fig:invariance} \emph{Gauge Invariance Postulate}.
            Along the top, a system in a state represented by~$(p_i; \var_i)$ evolves under
            map~$\map{M}$ to state represented by~$(p_i'; \var_i')$.  Along the
            bottom, a system in a state represented by~$(p_i; \var_i+
            \var_0)$ evolves under map~$\map{M}$ to a state represented~$(p_i'';
            \var_i'')$.  The postulate asserts that, for any~$\var_0$, one has~$p_i' = p_i''$
            for~$i=1, \dots, N$.}
\end{figure}

Second, we postulate that the result probabilities calculated from state~$\lvect{Q}' = \rmatrix{M}\lvect{Q}$ are unaffected by the above gauge transformation of the~$\var_i$ in~$\lvect{Q}$~(see Fig.~\ref{fig:invariance}). 
Remarkably, one finds that every orthogonal transformation,~$\rmatrix{M}$, of~$\lvect{Q}$ that satisfies this postulate can be rewritten as a unique unitary or antiunitary transformation of
\begin{equation}
\cvect{v} = (Q_1 + iQ_2, \dots, Q_{2N-1} + iQ_{2N})^\text{\textsf{T}}.
\end{equation}
Conversely, one finds that every unitary or antiunitary transformation of~$\cvect{v}$ is equivalent to a unique orthogonal transformation of~$\lvect{Q}$ which satisfies the global invariance postulate. That is, the set of all orthogonal transformations of~$\hypersphere$ is in one-to-one correspondence with the set of all unitary and antiunitary transformations of the space of unit vectors~$\cvect{v}$.  Hence, any physical transformation of the system in state~$\cvect{v}$ can be represented by a unitary or antiunitary transformation of~$\cvect{v}$.

\subsection*{Part 3:~Representation of Measurements}

\begin{figure}[!h]
                \includegraphics[width=3.4in]{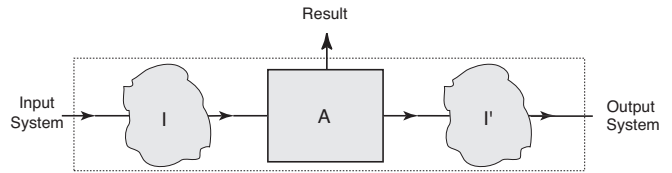}
                \caption{\label{fig:representation_of_measurements} \emph{Measurement Simulability 
                											Postulate}.
                An arrangement showing measurement~$\mment{A}$ flanked by interactions~$\interaction{I}$ 	
                and~$\interaction{I}'$.  The arrangement simulates
                measurement~$\mment{A}'$ insofar as its measurement probabilities
                and corresponding possible output states are concerned.}
\end{figure}

Generalizing from the fact that one can simulate any Stern-Gerlach measurement in terms any given Stern-Gerlach measurement flanked by suitable magnetic fields, we postulate that any reproducible measurement,~$\mment{A}'$, describable in the formalism can be simulated by an arrangement~(see Fig.~\ref{fig:representation_of_measurements}) where (a)~an interaction,~$\interaction{I}$, represented by a unitary transformation,~$\cmatrix{U}$,  acts on a system in some input state, (b)~the resulting system undergoes measurement~$\mment{A}$ which produces a measurement result and a system in some outgoing state, and (c)~some interaction,~$\interaction{I}'$, represented by a unitary transformation,~$\cmatrix{V}$, then acts upon the system to generate a system in some output state. 
From the reproducibility of~$\mment{A}$ and~$\mment{A}'$, one immediately obtains the Born rule.

\subsection*{Part 4:~Composite Systems}

On the basis of the above-mentioned classical-quantum correspondence argument, we are led to the assumption that the state of a system of definite energy,~$E$, whose observable degrees of freedom are time-independent, and in a time-independent background, evolves as~$(p_i; \var_i) \rightarrow (p_i; \var_i - E\Delta t/\alpha)$ during the interval~$[t, t+\Delta t]$, where~$\alpha\neq 0$ is a constant.  

Using this assumption and the above-mentioned global gauge condition, we obtain that, if a system consists of two subsystems in states~$(p_i; \var_i)$ and~$(p_j'; \var_j')$, respectively, where~$i=1, \dots, N$ and~$j=1, \dots, N'$, then the system has state~$(p_l''; \var_l'') = (p_i p'_j; \var_i + \var'_j)$, where~$l=N'(i-1) + j$.  The tensor product rule follows immediately.

\subsection*{Summary of Further Steps}

In Paper~II, a second classical-quantum correspondence argument is given that leads to the Average-Value Correspondence Principle which, in conjunction with the above assumption concerning the temporal evolution of a system, yields the explicit form of the unitary operator,~$\cmatrix{U}_t(dt) = \exp(-i\cmatrix{H}dt/\hbar)$, that represents temporal evolution of a system during the interval~$[t, t+dt]$ in terms of the Hamiltonian operator,~$\cmatrix{H}$.  Using the Average-Value Correspondence Principle, we then obtain the principal correspondence rules of quantum theory.


\section{Experimental Set-up and Postulates}
\label{sec:F}

\subsection{Abstract Experimental Set-up}
\label{sec:F-idealised-set-up}

A model of a physical system is a theoretical structure that allows one to make predictions about the behavior of the system in one of a set of possible experimental arrangements, where an experimental arrangement is abstracted as consisting of a preparation of the system, followed by an interaction and a measurement.  This set of experimental arrangements, which we shall refer to as an \emph{experimental set}, is a subset of all conceivable experimental arrangements in which the system could be placed.

Since the primary goal of the present work is to illuminate the physical origin of the quantum formalism, it is important to clearly delineate, at the outset, the experimental set to which the theoretical model, to be developed, is intended to apply.  
Below, we develop an operational procedure to delineate the experimental set, which is based on the following ideas. 

Consider an experimental arrangement where, in each run, a system~(taken from a source of identical systems) undergoes a preparation and an interaction, and is then subject to a measurement. Given that one's goal is to probe the behavior of the system, one ideally wishes to prepare the system in such a way that the data obtained from the measurement is \emph{independent} of arbitrary interactions with the system prior to the preparation.  In this way, one ensures that the measurement data are not influenced by conditions which are not under experimental control.  We shall say that experimental set-ups of this kind are \emph{closed}~(or have the property of \emph{closure}), and we shall restrict our consideration to such set-ups.   Using the concept of closure, we can then give a systematic procedure for generating the set of all closed experimental arrangements which, roughly speaking, probe the same \emph{behavioral aspect} of a system as does some given measurement~$\mment{A}$.  Such a set will be said to be an experimental set \emph{generated} by measurement~$\mment{A}$.

For example, any experimental arrangement where silver atoms are subject to preparation by a Stern-Gerlach device, undergo an interaction with a uniform magnetic field, and finally undergo a Stern-Gerlach measurement, is closed in the above sense.  Furthermore, the set of all such arrangements is an experimental set generated by any given Stern-Gerlach measurement. All such set-ups probe the same behavioral aspect of the system, namely its spin behavior.

Before we can precisely define closure and give a procedure for generating an experimental set, it is necessary to formulate a number of fundamental background assumptions and idealizations, to which we now turn.

\subsubsection{Background assumptions and Idealizations}

The theoretical framework of classical physics makes the following
key background assumptions:
\begin{itemize}
\item[(a)]\emph{Partitioning}. The universe is partitioned into a
system, the background environment~(or simply,
the~\emph{background})~\footnote%
{The background environment of a systems is, by definition, that
part of the environment of a system which non-trivially influences
the behavior of the system, but which is not reciprocally affected
by the system.  For example, if a planet in the gravitational
field of a star is modeled as a test particle in a fixed
gravitational field of the star, then the planet~(test particle)
is the system, and the gravitational field is its background.  If
a part of the environment is reciprocally affected by the system,
the system is enlarged to include this part of the environment.
For example, if the reciprocal affect of the planet on the star is
relevant, the system is enlarged to include the star, and the star
and planet are regarded as interacting subsystems within the
enlarged system.}
of the system, measuring apparatuses, and the rest of the
universe.
\item[(b)]\emph{Time}.  In a given frame of reference, one can
speak of a physical time which is common to the system and its
background, and which is represented by a real-valued
parameter,~$t$.
\item[(c)]\emph{States.} At any time, the system is in a definite
physical state, whose mathematical description is called the
mathematical state, or simply the \emph{state}, of the system. The
state space of the system is the set of all possible states of the
system.
\end{itemize}
Since these assumptions are not in obvious conflict with quantum
phenomena, we adopt them unchanged.
The classical framework also makes two key idealizations concerning
measurements:
        \begin{itemize}
        \item[A1] \emph{Operational Determinism.} The result of a
        measurement performed on a system is determined by
        experimentally-controllable variables.
        \item[A2] \emph{Continuum.} The values of the possible results
        of a measurement form a real-valued continuum.
        \end{itemize}
However, the experimental investigation of elementary quantum
phenomena, such as Stern-Gerlach measurements performed on silver
atoms, leads to the following reasonable modification of these
idealizations:
        \begin{itemize}
            \item[A1$'$] \emph{Statistical Operational Determinism.} The
            data obtained when a measurement is performed on a system
            are best modeled by a probabilistic source~\footnote{A probabilistic source
		is a black box which, upon each interrogation, yields one of a given
number of results with a given probability.} whose 
            probabilities are determined by experimentally-controllable
            variables.
            \item[A2$'$] \emph{Finiteness.} A measurement performed on a system
            has a finite number of possible results.
        \end{itemize}

\begin{figure*}
\includegraphics[width=6.75in]{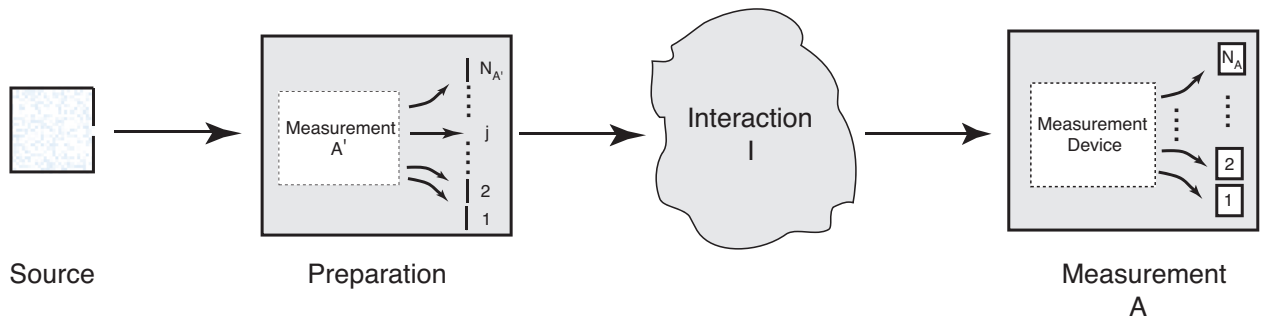}
\caption{\label{fig:idealised-set-up} An abstract experimental set-up.  In each run of the experiment, a physical system~(such as a
silver atom) is emitted from a source, passes a preparation step,
and is then subject to a measurement.   An
interaction,~$\interaction{I}$, may occur as indicated between the
preparation and measurement.  The preparation is
implemented as a measurement,~$\mment{A}'$, which has~$N_{A'}$
possible results, followed by the selection of those systems which
yield some result~$j$~($j=1,2, \dots, N_{A'}$). The
measurement,~$\mment{A}$, has~$N_{A}$ possible results. }
\end{figure*}

Using these assumptions as a basis, the general abstract
experimental set-up that we shall consider is shown
in~Fig.~\ref{fig:idealised-set-up}. A \emph{source} provides
identical copies of a physical system of interest. A
\emph{preparation step} either selects or rejects the incoming
system. In a particular run of the experiment, a physical system
from the source passes the preparation, and is then subject to a
\emph{measurement} or \emph{measurements}.

A measurement is idealized as a process that acts upon an
\emph{input} system, and generates an \emph{output} system together
with an observed \emph{result}.  The measurement detectors are
assumed not to absorb the systems that they detect. Accordingly, a
preparation consists of a measurement that determines to which
result the incoming system belongs, followed by the selection of
the system if the measurement registers a given result, and the
rejection of the system otherwise~\footnote{If detectors that do not absorb
the detected systems are unavailable, the preparation can instead be
implemented using a measurement where one of the detectors is
removed.  In this case, the detection of a system by the
subsequent measurement allows one to conclude that a system has
passed through the experimental apparatus.}.   In addition,
following the preparation, the system may undergo an
\emph{interaction} with a physical apparatus.

We shall only consider set-ups which satisfy particular
idealizations. In particular, we shall restrict consideration to
measurements that have the following properties:
\begin{itemize}
\item \emph{Distinctness.} The possible
results of a measurement have distinct values.

\item \emph{Reproducibility.} When a measurement is
immediately repeated, the same result is observed with certainty.

\item \emph{Classicality.} The measurements do not involve
auxiliary quantum systems.

\end{itemize}
The distinctness assumption excludes, at this stage, the
consideration of measurements where different results cannot be
observationally distinguished~\footnote{At the final stage of the
derivation~(Sec.~\ref{sec:non-distinctness}), we show how
measurements not satisfying the assumption of distinctness can be
theoretically described.}. The assumption of reproducibility is
drawn from classical physics and is adopted unchanged since it is
also a reasonable idealization in many quantum
experiments. Finally, the assumption of classicality ensures that
the relevant degrees of freedom of the measurement device itself can
be treated classically~\footnote{Once the quantum formalism is
obtained, the assumption of classicality can of course be relaxed to
allow measurements involving ancillary quantum systems to be
described.}.

In addition, we shall restrict consideration to  interactions that have the following
properties:
\begin{itemize}
\item \emph{Integrity-preserving.} The interactions preserve the
integrity of the system.

\item \emph{Reversible and Deterministic.} The interactions are
reversible and deterministic at the level of the state of the
system, and so can be represented as one-to-one maps over state
space.
\end{itemize}
The integrity-preserving assumption ensures that a theoretical model can follow a system during
the interaction.  The reversible and deterministic assumption is
drawn from classical physics, and is adopted unchanged.

We shall also assume that the background of the system can be
adequately modeled by classical physics insofar as its internal
dynamics is concerned.  For example, in the case of a system in a
background electromagnetic field, the field is assumed to be modeled
classically.  Similarly, we shall assume that parameters which
determine the measurement being performed~(the orientation of a
Stern-Gerlach apparatus, for instance) are described classically as
real numbers.  In short, it is assumed that the
non-classicality is entirely concentrated in the system and in its
interactions with the background and the measurement devices.

\subsubsection{Closed set-ups in the quantum framework}

From assumptions~A1$'$ and~A2$'$, it follows that, in a given
experimental set-up, the measurement data obtained in repeated
trials are theoretically characterized by a finite set of
probabilities.   Therefore, the closure condition defined earlier,
when applied in the context of these assumptions, requires that
these probabilities are independent of the pre-preparation
history of the system.

\subsubsection{Experimental sets}

The measurements employed in the abstract set-up are chosen from a
\emph{measurement set},~$\set{A}$, which we shall define below. As
mentioned previously, it will be assumed that each measurement has
the property of finiteness, which we shall now operationalize by
saying that, when the measurement is carried out on a system which
has been emitted from the source and has undergone arbitrary
interactions thereafter~%
\footnote{Here and subsequently, it is assumed that all interactions
with the system preserve the integrity of the system.},
the measurement generates one of a finite
number of \emph{possible results}, a possible result being defined
as one that has a non-zero probability of occurrence.

Consider now an experiment~(Fig.~\ref{fig:idealised-set-up}) in
which a system from a source is subject to a preparation consisting
of measurement,~$\mment{A'}$, with~$N_{A'}$ possible results, with
result~$j$ selected~($j=1,\dots, N_{A'}$), followed by
measurement~$\mment{A}$~(with~$N_A$ possible results),
\emph{without} an interaction in the intervening time.

Suppose that the data obtained in the experiment are characterized by a 
probabilistic source with~$N_A$ possible
results and with probability n-tuple~$\vect{p} = (p_1, p_2, \ldots, p_{N_A})$,
where~$p_i$ is the probability of the~$i$th result~$(i=1, 2, \dots,
N_A)$. 
 If, for all~$j$, $\vect{p}$ is
 independent of arbitrary pre-preparation interactions with the system, the
preparation will be said to be \emph{complete} with respect to
 measurement~$\mment{A}$.  If this completeness condition also holds
true when~$\mment{A}$ and~$\mment{A}'$ are interchanged,
then~$\mment{A}$ and~$\mment{A}'$ will be said to form a
\emph{measurement pair}.

The set of measurements \emph{generated} by~$\mment{A}$ forms a
measurement set,~$\set{A}$, which is defined as the set of all
measurements that~(i) form a measurement pair with~$\mment{A}$ and
that~(ii) are not a composite of other measurements in~$\set{A}$. An
important corollary of this definition is that two measurement sets
are either identical or disjoint.

Interactions following the preparation step are chosen from
an \emph{interaction set},~$\set{I}$, which is defined as follows.
Suppose that, in the experiment of Fig.~\ref{fig:idealised-set-up},
an interaction,~$\interaction{I}$, occurs between the preparation
and measurement. If, for all~$\mment{A},\mment{A}' \in \set{A}$, the
preparation remains complete with respect to the subsequent
measurement, then~$\interaction{I}$ will be said to be
\emph{compatible} with~$\set{A}$. The set~$\set{I}$
is then defined as the set of all such compatible interactions.

In terms of these definitions, a closed set-up consists of a source
of systems where each system is prepared using a
measurement~$\mment{A}' \in\set{A}$, is then subject to an
interaction~$\interaction{I}\in \set{I}$, and then undergoes a
measurement~$\mment{A} \in\set{A}$, where~$\set{A}$ is generated
by~$\mment{A}$, and~$\set{I}$ is the set of all interactions
compatible with~$\set{A}$.  The set of all such set-ups will be taken 
as constituting an experimental set, and will be said to be \emph{generated} by measurement~$\mment{A}$.

Finally, if there are two experimental set-ups, each with a source
containing identical copies of the same physical system, constructed using measurements from measurement sets~$\set{A}^{(1)}$ and~$\set{A}^{(2)}$, respectively, then, if the measurement sets are disjoint, then the two set-ups will be said to be \emph{disjoint}.

\subsubsection{An example}

To illustrate the above definitions, consider an experiment where
silver atoms emerge from a source~(an evaporator), pass through a
Stern-Gerlach preparation device, undergo an interaction, and
finally undergo a Stern-Gerlach measurement. In this case, one finds
experimentally that the set,~$\set{A}$, generated by any
Stern-Gerlach measurement consists of all Stern-Gerlach measurements
of the form~$\mment{A}_{\theta, \phi}$, where~$(\theta, \phi)$ is
the orientation of the Stern-Gerlach device.  Measurements that are
composed of two or more Stern-Gerlach measurements are excluded
from~$\set{A}$ by definition.

Consider now an interaction,~$\interaction{I}_{\theta_B, \phi_B, t,
\Delta t}$, consisting of a uniform $\vect{B}$-field acting during
the interval~$[t,t+\Delta t]$ in some direction~$(\theta_B,
\phi_B)$.  If such an interaction occurs between the preparation and
measurement, one finds experimentally that the completeness of the
preparation with respect to the measurement is preserved; that is,
the interaction is compatible with~$\set{A}$. Hence,
all interactions in which a uniform magnetic field acts between the
preparation and measurement are in the interaction set,~$\set{I}$.

The experimental set in this case consists of all set-ups consisting of
a Stern-Gerlach preparation, an interaction with a uniform magnetic field, followed by
a Stern-Gerlach measurement.  

Finally, to illustrate the concept of disjoint set-ups, consider a
source which emits a system consisting of two distinguishable
spin-1/2 particles on each run of an experiment, and consider two
set-ups where the first set-up is constructed using measurement set~$\set{A}^{(1)}$
consisting of all possible Stern-Gerlach measurements performed on
one of the particles, and the second is constructed using a measurement
set~$\set{A}^{(2)}$ consisting of all possible Stern-Gerlach
measurements performed on the other particle. In this case,
according to the above definitions, it is found that the two
measurement sets are disjoint. The set-ups themselves are
accordingly said to be disjoint.

\subsubsection{Some remarks}

The above definitions can be understood intuitively as follows. If
the measurements~$\mment{A}$ and~$\mment{A}'$ form a measurement
pair, then they can be regarded as probing the same behavioral aspect of a system. 
For example, if measurements~$\mment{A}$ and~$\mment{A}'$ are
Stern-Gerlach measurements, in which case~$\mment{A}$
and~$\mment{A}'$ form a measurement pair, both are probing the
spin behavior of a system.  A measurement set can then be understood
as consisting of \emph{all} measurements that probe a given
behavioral aspect of a system.

An interaction which is compatible with a measurement set can, similarly, be understood as one that does not allow the
behavioral aspect that is being probed to be influenced by
degrees of freedom that belong to some other behavioral aspect. For
example, the effect on a silver atom of a uniform magnetic field is only dependent upon the spin
degrees of freedom of the system, and is therefore compatible with a
measurement set consisting of Stern-Gerlach measurements. However,
an interaction that affects the spin degrees of freedom but is
dependent upon the spatial degrees of freedom of the system is not
compatible with this measurement set.

Finally, if two set-ups are disjoint, they are probing distinct
behavioral aspects of the same physical system.  For example, a
system may, in one set-up, be subject to measurements from a
measurement set consisting of Stern-Gerlach measurements, and, in
another set-up, to measurements from a measurement set that probe
the spatial behavior of the system.  As we would expect from
classical physics, these measurement sets are disjoint, which can be
understood as reflecting the fact that these set-ups are examining
disjoint behavioral aspects~(spin and spatial behavior) of the
system.

\subsection{Statement of the Postulates}
\label{sec:F-statement-of-postulates}

Consider the idealized experiment illustrated in
Fig.~\ref{fig:idealised-set-up} in which a system passes a
preparation step that employs a measurement~$\mment{A}'$ in
measurement set~$\set{A}$, undergoes an
interaction,~$\interaction{I}$ in the interaction set~$\set{I}$, and
is then subject to a measurement,~$\mment{A} \in \set{A}$,
where~$\set{A}$ is generated by~$\mment{A}$, and~$\set{I}$ is the set
of all interactions compatible with~$\set{A}$.  The abstract theoretical model that describes this set-up
satisfies the following postulates.

\begin{enumerate}
\item[1.] \textbf{Measurements}
    \begin{itemize}
   \item[1.1]
        \emph{Results.}
            When any measurement~$\mment{A} \in \set{A}$ is
            performed, one of~$N$~($N\geq 2$) possible results are
            observed.

    \item[1.2]
        \emph{Measurement Simulability.}
            For any given pair of measurements~$\mment{A},\mment{A}' \in \set{A}$, there
            exist interactions~$\interaction{I}, \interaction{I}' \in \set{I}$
            such that~$\mment{A}'$ can, insofar as probabilities of
            the observed results
            and insofar as the possible output states of the measurement are concerned, be simulated 
            by an arrangement where~$\interaction{I}$ is immediately
            followed by~$\mment{A}$ which, in turn, is immediately
            followed by~$\interaction{I}'$~(see
            Fig.~\ref{fig:representation_of_measurements}).
    \end{itemize}
  \item[2.] \textbf{States}
    \begin{itemize}
    \item[2.1]
        \emph{Complementarity.}
    When any given measurement~$\mment{A} \in \set{A}$ is performed on the system, one of $2N$~possible outcomes is realized with probability~$P_1, \dots, P_{2N}$, respectively.  The individual outcomes are unobserved.  Result~$i$ is observed whenever outcome~$(2i-1)$ or outcome~$2i$ is realized~(see Fig.~\ref{fig:prob-tree-complementarity}).  
    
    \item[2.2] \emph{States.}  The state,~$\state{S}$, of the system with respect to measurement~$\mment{A} \in\set{A}$ is given by~$\lvect{Q} = (Q_1, Q_2, \dots, Q_{2N})\transpose$, where~$Q_q \in [-1,1]$, $q=1, \dots, 2N$.  The probability of outcome~$q$ is given by~$P_q = Q_q^2$, and the variable,~$\sigma_q = \operatorname{sign}(Q_q)$, which is defined if~$P_q \neq 0$, is a binary degree of freedom~(a \emph{polarity}) associated with outcome~$q$ and is physically realized whenever outcome~$q$ is realized, but is unobserved.
    
    \item[2.3]
        \emph{State Representation.}
    The state,~$\state{S}$, of a system with respect to measurement~$\mment{A}$ can be represented by the pair~$(\vect{p}, \vect{\var})$ where~$\vect{p}=(p_1, \dots, p_N)$ and~$\vect{\var}=(\var_1, \dots, \var_N)$ are real $n$-tuples, and where~$p_i$ is the probability that the~$i$th result of measurement~$\mment{A}$ is observed.  In particular, the state is given by~$\lvect{Q} = (\sqrt{p_1} q_{a|1}, \sqrt{p_1} q_{b|1}, \dots, \sqrt{p_N} q_{a|N}, \sqrt{p_N} q_{b|N})$, where~$q_{a|i} = f(\var_i)$ and~$q_{b|i} = \tilde{f}(\var_i)$, where~$f$ is not a constant function, and where~$f$ and~$\tilde{f}$ are differentiable functions~(see Fig.~\ref{fig:prob-tree}).

    \item[2.4]
        \emph{Information Metric.}
            The metric over~$\lvect{P} =  (P_1, P_2, \dots, P_{2N-1}, P_{2N})$ is the information metric,~$ds^2 =\frac{1}{4} \sum_{q=1}^{2N} dP_q^2/P_q^2.$

    \item[2.5]
        \emph{Measure Invariance.}
            The measure,~$\mu(p_1, \dots, p_N; \var_1, \dots, \var_N)$, over~$\vect{p}, \vect{\var}$  induced by the metric over~$\lvect{Q}$ satisfies the condition~$\mu(p_1, \dots, p_N; \var_1, \dots, \var_N) = \mu(p_1, \dots, p_N; \var_1 + \var_0, \dots, \var_N + \var_0)$ for all~$\var_0$.
     \end{itemize}

\item[3.] \textbf{Transformations} 
    \begin{itemize}
    \item[3.1] \emph{Mappings.} Any physical transformation of the system,
         whether active~(due to temporal evolution) or passive~(due to a change of
         frame of reference),  is represented by a map,~$\map{M}$, over the state space,~$\spc{S}$, of the system.
         
    \item[3.2] \emph{One-to-one.} Every map,~$\map{M}$, that represents a physical transformation of the 		system is one-to-one.    
    
    \item[3.3] \emph{Continuity.} If a physical
    transformation is continuously dependent upon the real-valued
    parameter n-tuple~$\Bpi$, and is represented by the
    map~$\map{M}_{\Bpi}$, then~$\map{M}_{\Bpi}$ is continuously dependent
    upon~$\Bpi$.

    \item[3.4] \emph{Continuous Transformations.}  If~$\map{M}_{\Bpi}$
    represents a continuous transformation, then, for some value
    of~$\Bpi$,~$\map{M}_{\Bpi}$ reduces to the identity.

    \item[3.5] \emph{Metric Preservation.}  The map~$\map{M}$ preserves the metric
    over the state space,~$\spc{S}$, of the system.

    \item[3.6] \emph{Gauge Invariance.} The map~$\map{M}$ is such that,
    for any state~$\state{S} \in \spc{S}$, the
    probabilities,~$p_1', p_2', \dots, p_N'$,
    of the results of measurement~$\mment{A} \in \set{A}$ performed upon a system
    in state~$\state{S}'=\map{M}(\state{S})$ are unaffected if, in any
    representation,~$(p_i; \var_i) \equiv (\vect{p}, \vect{\var})$, of the
    state~$\state{S}$ written down with respect to~$\mment{A}$,
    an arbitrary real constant,~$\var_0$, is added
    to each of the~$\var_i$~(see Fig.~\ref{fig:invariance}).

    \item[3.7] \emph{Temporal Evolution.} The map,~$\map{M}_{t, \Delta t}$,
    which represents temporal evolution of a system in a time-independent
    background during the interval $[t, t+\Delta t]$, is such that any
    state,~$\state{S}$, represented as~$(p_i; \var_i)$, of
    definite energy~$E$,  whose observable degrees of freedom are
    time-independent, evolves to~$(p_i'; \var_i') =
    (p_i; \var_i - E\Delta t/\alpha)$, where~$\alpha$
    is a non-zero constant with the dimensions of action.
    \end{itemize}

\end{enumerate}

The above postulates, together with the Average-Value Correspondence
Principle~(AVCP), which will be given in Paper~II, suffice to
determine the form of the abstract quantum model for the abstract
set-up. From the Results postulate, it follows that, when any
measurement in~$\set{A}$ is performed on the system, one of~$N$
possible results is observed. Accordingly, we shall denote the
abstract quantum model of such a set-up by~$\model{q}(N)$.

Finally, we shall need the Composite Systems postulate, below, in order to obtain a
rule, which we shall refer to as the \emph{composite systems rule},
for relating the quantum model of a composite system to the quantum
models of its component systems:
\begin{enumerate}
\item[4.] \textbf{Composite Systems}. Suppose that a system admits
a quantum model with respect to the measurement
set~$\set{A}^{(1)}$ whose measurements have~$N$ possible
results, and admits a quantum model with respect to
measurement set~$\set{A}^{(2)}$ whose measurements have~$N'$
possible results, where the sets~$\set{A}^{(1)}$
and~$\set{A}^{(2)}$ are disjoint.

Consider the quantum model of the system with respect to the
measurement set~$\set{A} = \set{A}^{(1)} \times \set{A}^{(2)}$
that contains all possible composite measurements consisting of a
measurement from~$\set{A}^{(1)}$ and a measurement
from~$\set{A}^{(2)}$. If the states of the subsystems can be
represented as~$(p_i; \var_i)$~$(i=1, 2, \dots,
N)$ and~$(p_j'; \var_j')$~$(j=1, 2, \dots,
N')$, respectively, then the state of the composite system
can be represented as~$(p_l''; \var_l'')$~$(l=1, 2, \dots,
NN')$,
where~$p_l''=p_i p_j'$ and~$\var_l'' = \var_i +
\var_j'$, where~$l=N'(i-1) + j$.
\end{enumerate}

\section{Physical comprehensibility of the Postulates}
\label{sec:overview}

When formulating the postulates, our goal has been to maximize their
physical comprehensibility. For the purposes of discussion, it is
helpful to distinguish two \emph{levels} of physical
comprehensibility. First, at the minimum, a comprehensible postulate
is one that can be transparently understood as a simple assertion
about the physical world.  If this is the case, we shall say that
the postulate has the property of \emph{transparency}. Second, a
postulate has an additional level of comprehensibility if it can
also be traced to well-established experimental facts and physical
ideas or principles, a property we shall refer to as \emph{traceability}.

To illustrate these ideas, consider the example of Einstein's
postulate that the speed of light emitted by a source is independent of the speed of the source. The postulate can be transparently understood as a physical assertion in itself.  In addition, the postulate can also be
understood as a reasonable inference from the well-established
results~(namely, the constancy of the two-way speed of light) of the Michelson-Morley experiment, the generalization from the specific context of the experiment being achieved by an appeal to the general principle of the uniformity of nature. Hence, the postulate is both transparent and traceable.

In our discussion below, we shall organize the postulates into three groups according to their origin:
\begin{itemize}
\item[1.]  \textbf{Based on Experimental Observations.}
Postulates obtained through direct generalization of experimental
facts that are taken to be characteristic of quantum phenomena.

\item[2.]  \textbf{Drawn from Classical Physics.}
    \begin{itemize}
    \item[2.1] Postulates adopted unchanged from the theoretical
    framework of classical physics.
    \item[2.2] Postulates obtained from classical physics via a
    classical--quantum correspondence argument.
    \end{itemize}

\item[3.]  \textbf{Novel Postulates.}  Postulates based on
novel theoretical principles or ideas which cannot obviously be
traced to classical physics or to experimental observations.

\end{itemize}
The first group of postulates are obtained by direct generalization of elementary 
experimental facts, such as the statistical nature of experimental results, that can be reasonably taken as characteristic of quantum phenomena.  Insofar as these postulates can be regarded as reasonable generalizations of experimental facts, they can be regarded as possessing transparency and traceability

In formulating the second group of postulates, we recognize that the assumptions underlying the theoretical framework of classical physics are transparent
and traceable to well-established experimental facts and theoretical
ideas, and remain fundamental to the way in
which we conceptualize the physical world.  Accordingly, we
attempt to preserve these assumptions as far as possible in the face of quantum phenomena.
In particular, some of the postulates are obtained by simply
adopting fundamental features of the classical theoretical framework, while the others
are obtained by transposing particular features of the classical
models of physical systems into the quantum realm via a
classical--quantum correspondence argument. 

Finally, the third group of postulates consist of novel postulates, one physical postulate~(the Complementarity postulate) and two information-geometric postulates~(the Information Metric and the Metric Preservation postulates).

\subsection{Postulates based upon experimental facts}

\subsubsection*{Postulate~1.1: Results}

Consider an experiment in which Stern-Gerlach preparations and
measurements are performed upon silver atoms, and where the
set~$\set{A}$ consists of the elements~$\mment{A}_{\theta, \phi}$
representing Stern-Gerlach measurements in the direction~$(\theta,
\phi)$.  In this experimental set-up, which is closed in the sense defined earlier, we find
that each measurement yields one of two possible results.  The Results
postulate generalizes this finding by asserting that all the measurements in
a measurement set have the same number,~$N$, of possible results.

\subsubsection*{Postulate~1.2: Measurement Simulability}

Consider again the above Stern-Gerlach experiment. In this experiment, 
if an interaction consisting of a
uniform magnetic field acts between the preparation and measurement,
one finds that the probabilities of the measurement results are the
same as those obtained if a different Stern-Gerlach measurement is
performed with the interaction absent.

Using this observation, one finds that it is possible to simulate
measurement~$\mment{A}_{\theta, \phi}$ using any given
measurement~$\tilde{\mment{A}} \in \set{A}$ if followed
immediately before and after by suitable interactions.  The
simulation behaves precisely as~$\mment{A}_{\theta, \phi}$
insofar as the probabilities of measurement results~$1$ and~$2$,
and the corresponding output states, are concerned. Postulate~1.2
can be regarded as direct generalization of this observation.

\subsection{Postulates adopted unchanged from classical physics}

A classical model of a physical system is based upon the
\emph{partitioning}, \emph{time} and \emph{states} background
assumptions given earlier, and these are adopted unchanged in the
abstract quantum model.  In addition to the assumptions~$A1$ and~$A2$ 
concerning measurements given earlier, the classical model additionally makes the
following key assumptions:
\begin{enumerate}

    \item[B] \emph{States.}

        \begin{itemize}
        \item[B1] \emph{Determinism.} The state of the system and
        a theoretical description of a measurement that is performed on
        the system determine the measurement result.
        \end{itemize}

    \item[C] \emph{Transformations.}
        \begin{itemize}
        \item[C1] \emph{Mappings.} Physical transformations of the system,
         either due to temporal evolution or due to a passive change of
         frame of reference,  are represented by maps over the space
          of states.
        \item[C2] \emph{One-to-one.} The mappings are one-to-one.
        \item[C3] \emph{Continuity.} If a map represents
        a physical transformation that depends continuously upon a
        real-valued set of parameters, then the map is continuously
        dependent upon these parameters.
        \item[C4] \emph{Continuous transformations.} If a map represents
        a continuous transformation~(such as temporal evolution) that
        depends continuously upon a set of real-valued parameters, then,
        for some value of these parameters, the map reduces to the
        identity.
        \end{itemize}
\end{enumerate}

We remark that, in asserting~C1--C2, it is presupposed that
physical transformations of a physical system are deterministic and
reversible, which prevents the description of irreversible or
indeterministic transformations within the classical framework at a
fundamental level.

First, we consider those postulates which adopt classical
assumptions unchanged.  The Mappings and One-to-one postulates respectively correspond
to assumptions~C1 and~C2, while the Continuity and Continuous Transformations postulates correspond to  assumptions~C3 and~C4, respectively.

Second, as described earlier, in light of the results of experiments
involving quantum systems~(such as Stern-Gerlach measurements on
silver atoms), it is reasonable to modify assumptions~A1,~A2 to
assumptions~A1$'$ and~A2$'$~(see Sec.~\ref{sec:F-idealised-set-up}),
and accordingly to modify~B1 as follows:
    \begin{itemize}
    \item[B1$'$] \emph{Statistical Determinism.} The state of the
    system and a theoretical description of a measurement that is
    performed on the system only \emph{statistically} determine
    the measurement result.
    \end{itemize}
Assumption~B1$'$ is incorporated within the Complementarity postulate.

\subsection{Postulates obtained through classical-quantum
correspondence}

A general guiding principle in building up a quantum model of a
physical system is that, in an appropriate limit, the predictions of
the quantum model of the system stand in some one-to-one
correspondence with those of a classical model of the system.   By
establishing such a correspondence between the quantum and classical
models of a particle, we shall transpose two elementary
properties of the classical model across to the quantum model and
then, by generalization, transpose these properties across to the abstract quantum
model,~$\model{q}(N)$.  In this manner, we shall obtain a number of important postulates.

The key idea, explicated in detail below, is that a quantum model of a particle moving in one dimension corresponds, in the appropriate classical limit, to the classical ensemble model of the particle, namely the Hamilton-Jacobi ensemble model.  We consider coarse position measurements, and discretize the Hamilton-Jacobi state~$(P(\vect{r},t), S(\vect{r},t))$, where~$P(\vect{r},t)$ is the probability density function over position and~$S(\vect{r},t)$ is the action function, as~$(p_i^{(\text{CM})}; S_i)$, where~$p_i^{(\text{CM})}$ is the probability that the position measurement yields result~$i$~($i=1, \dots, N)$, and~$S_i$ is the action associated with position~$i$.  On the assumption that the quantum state can be put into one-to-one correspondence with the classical state in the classical limit, it follows that the quantum state of the particle must be of the form~$(p_i; \var_i)$, where, again, the~$p_i$ are the probabilities of the position measurement, and where the~$\var_i$ are $N$~real degrees of freedom.  In the limit, we assume~$p_i = p_i^{(\text{CM})}$ and~$S_i = \alpha\var_i$, where~$\alpha$ has the dimensions of action.  

Using this correspondence, we can transpose elementary features of the Hamilton-Jacobi model across to the quantum model of the particle.  Two properties will be of importance, to which we shall refer as \emph{global gauge invariance} and \emph{temporal evolution}. 

The first property arises from the fact that the transformation~$(p_i^{(\text{CM})}; S_i) \rightarrow (p_i^{(\text{CM})}; S_i + S_0)$, where~$S_0$ is an arbitrary real number, is a global gauge transformation, which therefore leaves the predictions of the classical model invariant.  This leads to the assumption that, in the quantum model of the particle, the transformation~$(p_i; \var_i) \rightarrow (p_i; \var_i + \var_0)$ is also a global gauge transformation.  We then generalize by assuming that this also holds of the abstract quantum model,~$\model{q}(N)$.  This global gauge invariance condition leads to two postulates.

First, from the global gauge invariance condition, it follows that, in particular, the map,~$\map{M}$, representing physical transformation of a system described using model~$\model{q}(N)$, which takes the state~$(p_i; \var_i)$ to~$(p_i'; \var_i')$, is such that the~$p'_i$ are invariant under the transformation~$(p_i; \var_i) \rightarrow (p_i; \var_i + \var_0)$ for any~$\var_0$.   This is the content of the Gauge Invariance postulate, and can be seen to impose a constraint upon~$\map{M}$.  

Second, we require that the measure,~$\mu(p_1, \dots, p_N; \var_1, \dots, \var_N)$,  induced by the metric over state space is consistent with the global gauge transformation, so that~$\mu(p_1, \dots, p_N; \var_1+\var_0, \dots, \var_N + \var_0) = \mu(p_1, \dots, p_N; \var_1, \dots, \var_N)$ for any~$\var_0$, which is the  content of the Measure Invariance postulate, and imposes a constraint upon the functions~$f$ and~$\tilde{f}$.

The second property arises by considering the special case of a classical ensemble in a time-independent background with state~$(p_i^{(\text{CM})}(t); S_i(t))$, whose observable degrees of freedom~(namely the~$p_i^{(\text{CM})}$ and the~$\Delta S_l \equiv S_{l+1} - S_l$, with~$l=1, \dots, N-1$) are time-independent.  In this case, the state evolves in time~$\Delta t$ to~$(p_i^{(\text{CM})}(t); S_i(t) -
E\Delta t)$, where~$E$ is the total energy of the system.  That is, in this particular case, the temporal rate of change of the~$S_i$ encodes the energy of the system.  Using the correspondence~$S_i = \alpha\var_i$, we assume that, for a system described within the model~$\model{q}(N)$ in a time-independent background, and in a state (i)~whose observable degrees of freedom are time-independent, and (ii)~which is of definite energy~$E$, the temporal rate of change of the~$\var_i$ is~$E/\alpha$.    This constitutes the Temporal Evolution postulate.

From the global gauge invariance condition and the Temporal Evolution postulate, we obtain a further postulate concerning composite systems  We consider  a composite system consisting of two subsystems and require that  (a)~a global gauge transformation on either subsystem leads only to a global gauge transformation on the composite system, and (b)~in the case where the observable degrees of freedom of both subsystems are time-independent, the subsystems are in states of definite energy~$E$ and~$E'$, respectively, and their environment is time-independent, the energy of the composite system is~$E'' = E + E'$.  From these two ideas, we obtain the Composite Systems postulate.

\subsubsection{The Correspondence Argument}

Consider an experiment in which a position measurement is used to
prepare a particle at time~$t_0$, and a position measurement is
subsequently performed at time~$t_1$, during which interval a
potential~$V(\vect{r},t)$ is assumed to act.  When such an
experiment is actually performed, one necessarily uses position
measurements with a finite number of possible measurement results. In this
case, the experimental results~(where, for instance, an electron
passes through a sub-micron aperture, is subject to electric-field
interactions, and is subsequently detected on a screen) support
the conclusion that, if these coarse position measurements are of
sufficiently high spatial resolution, the preparation is, to a
very good approximation, complete with respect to the subsequent
measurement.

Suppose, then, that a coarse position measurement with~$N$
possible results is used to implement both the preparation and
measurement steps, and further let us suppose that the coarse
measurement is such that the probability that a detection is
obtained in any run of the experiment is very close to unity.
Further, let us suppose that the coarse measurement is of
sufficient resolution that the preparation can be regarded as
being complete with respect to the measurement.  Then we can form
a quantum model, which we shall denote~$\model{q}^*(N)$, within
the framework of the abstract quantum model~$\model{q}(N)$, which
approximately describes the experiment after time~$t_0$.

By the Results postulate and the assumption~B1$'$ above, the
state,~$\state{S}(t_1)$, of the system immediately prior to the
coarse position measurement determines the probability
n-tuple,~$\vect{p}(t_1)=(p_1, \dots, p_N)$, where~$p_i$ is the
probability of detection at the $i$th detector, which characterizes
the data obtained from the coarse position measurement.

If the above experiment is repeated, except that the coarse
position measurement is delayed until time~$t_2$,
then~$\state{S}(t_1)$, together with a theoretical representation
of any interaction in the interval~$[t_1, t_2]$, must~(by
assumption~B1$'$) enable the prediction of the probability
n-tuple~$\vect{p}(t_2)$ that describes the coarse position
measurement data obtained at time~$t_2$.  To determine what
additional degrees of freedom the state~$\state{S}(t_1)$ must
contain in order to make this prediction possible, consider the
classical limit.

Suppose that~$m$ is increased towards values characteristic of
macroscopic bodies. Under the assumption made above, the
preparation is complete with respect to the measurement, so that
the system continues to be well-described by the
model~$\model{q}(N)$ even in this classical limit.
However, as~$m$ tends towards macroscopic values, it is reasonable
to expect that the system will increasingly behave in accordance
with its classical model between times~$t_1$ and~$t_2$. That is,
in this classical limit, we expect that~$\vect{p}(t_2)$, which is
determined in the quantum model in terms of~$\vect{p}(t_1)$ and
the other degrees of freedom in~$\state{S}(t_1)$, will coincide
with the n-tuple~$\vect{p}^{(\text{CM})}(t_2)$ that is predicted
by a classical model of a particle of mass~$m$ moving in the same
potential.

The relevant classical model in this situation is a particle
ensemble model.  For such an ensemble model, one can choose to
describe an ensemble for the case of given total energy by means
of a probability density function over phase space, and to
describe the evolution of this function using Newton's equations
of motion. Alternatively, one can employ the Hamilton-Jacobi
model, which is physically equivalent.  We choose the latter since
it is more easily described on a discrete spatial lattice.

In the Hamilton-Jacobi model, the state of the ensemble is given
by~$(P(\vect{r},t), S(\vect{r}, t))$, which satisfies the
Hamilton-Jacobi equations,
\begin{equation}\label{eqn:HJ}
\begin{gathered}
\frac{\partial P}{\partial t} + \nabla. \left(
                        \frac{1}{m} P \, \nabla  S
                                        \right) = 0 \\
\frac{1}{2m} \left( \nabla S \right)^2 + V(\vect{r}, t) = -
\frac{\partial S}{\partial t},
\end{gathered}
\end{equation}
where~$P(\vect{r},t)$ is the probability density function over position and~$S(\vect{r},t)$ is the action function.
In the case of coarse position measurements with~$N$ possible
results, we shall use the discretized form of the Hamilton-Jacobi
state,~$(p_1^{(\text{CM})}, \dots, p_N^{(\text{CM})}, S_1, \dots,
S_N) = (p_i^{(\text{CM})}; S_i)$, where~$p_i^{(\text{CM})}$ is the
probability that the position measurement yields a detection at
the~$i$th measurement location, and~$S_i$ is the classical action at
the~$i$th measurement location.

In order that the predictions of the quantum and classical models
agree in the classical limit, the quantum
state~$\state{S}(t)$~($t>t_0$) must contain degrees of freedom which
encode~$N$ quantities, which we shall denote~$S^{(\text{QM})}_1,
\dots, S^{(\text{QM})}_N$, which, in the classical limit, are equal
to the~$S_i$.  Equivalently, we shall assume that~$\state{S}$
contains~$N$ dimensionless real quantities,~$\var_1, \dots, \var_N$,
such that~$S^{(\text{QM})}_i = \alpha\var_i$, where~$\alpha$ is a
non-zero constant with the dimensions of action.

From the above discussion, in the model~$\model{q}^*(N)$, the
state,~$\state{S}$, is given by~$(\vect{p}, \vect{\var})$,
where~$\vect{\var}= (\var_1, \dots, \var_N)$. A direct generalization of this observation leads to the
assumption that the state of a system described by the abstract model~$\model{q}(N)$ with respect to some measurement~$\mment{A}$ can be represented by~$(\vect{p}, \vect{\var})$.  As will be discussed below, this assumption provides the motivation for the State Representation postulate.

\subsubsection{Global Gauge Invariance} 

In the continuum Hamilton-Jacobi model, the observables associated with~$S(x,t)$ for a system in state~$(P(x,t), S(x,t))$ are~$\partial S/\partial x$ and~$\partial S/\partial t$.  Hence, the transformation~$S(x,t) \rightarrow S(x,t) + S_0$ is a global gauge transformation of the model.   Therefore, the discretized form of the model has a global gauge invariance property, namely that, for a system with state~$(p^{(CM)}_i; S_i)$, the transformation~$S_i \rightarrow S_i + S_0$ for~$i=1, \dots, N$ and for any~$S_0$ is a global gauge transformation, leaving invariant all physical predictions made on the basis of the state.  
 
From this property of the Hamilton-Jacobi model, using the above classical-quantum correspondence, we assume, in the quantum model of a particle, and, even more generally for the abstract quantum model~$\model{q}(N)$, the transformation
\begin{equation}
(p_i; \var_i) \rightarrow (p_i; \var_i + \var_0),
\end{equation}
where~$\var_0 \in \numberfield{R}$, is also a gauge transformation.  From this assumption, we now shall draw two postulates.  

\subsubsection*{Postulate~3.6:~Gauge Invariance}  

First, we note that, as a direct result of this global gauge invariance assumption, it follows that a transformation~(representing passive or active physical transformation of the system) of the state~$(p_i, \var_i)$ to the
state~$(p_i'; \var_i')$ is such that the~$p_i'$ are unchanged if an arbitrary real constant,~$\var_0$, is added to each of the~$\var_i$.    This is the content of the Gauge Invariance postulate, which may be regarded as a specific example of the assumed global gauge invariance property.

\subsubsection*{Postulate:~2.5:~Measure Invariance} 

Second, we impose the requirement that the measure~(or, in the language of Bayesian probability theory, the \emph{prior}) over~$p_1, \dots, p_N, \var_1, \dots, \var_N$ induced by the metric over state space~(which metric arises from the Information Metric postulate) is compatible with the global gauge invariance property, and therefore satisfies the relation
\begin{equation} \label{eqn:measure-constraint}
\mu(p_i; \var_1, \dots, \var_N) = \mu(p_i; \var_1 + \var_0, \dots, \var_N +
\var_0),
\end{equation}
for any~$\var_0$, which is the content of the Measure Invariance postulate.

The requirement of the consistency of the measure with the global gauge invariance property can be understood as follows. Suppose that one is performing Bayesian inference on the quantum system.  If one's knowledge about the quantum system includes the fact that it has a global gauge invariance property, then the prior over the~$p_i$ and~$\var_i$ which one employs should reflect this fact.   Otherwise, one's inference will sometimes lead to predictions that are not consistent with  the global gauge invariance property.

\subsubsection*{Postulate~3.7:~Temporal Evolution}  Consider the special case of a system in a time-independent background whose observable degrees of freedom are time-independent.  According to the Hamilton-Jacobi equations, the state of such a system evolves in time~$\Delta t$ as
\[ (P(x,t+\Delta t), S(x, t+\Delta t)) = (P(x,t), S(x,t) - E\Delta t), \]
where~$E$ is the energy of the ensemble.   That is, the temporal rate of change of the unobservable degree of freedom encodes the total energy of the system.

Using the above classical-quantum correspondence, we assume that the quantum model of a particle in a time-independent background which is in a state of definite energy with time-independent observable degrees of freedom, evolves as~$(p_i; \var_i) \rightarrow (p_i; \var_i -E/\alpha\Delta t)$ during the interval~$[t, t+\Delta t]$.  The Temporal postulate directly transposes this assumption to the quantum model~$\model{q}(N)$.  

\subsubsection*{Postulate~4:~Composite Systems}

Let us consider a composite system, described in the model~$\model{q}(N'')$, consisting of two subsystems that are known to be in states represented by~$(p_i; \var_i)$ and~$(p'_j; \var'_j)$, respectively, with~$i=1, \dots, N$ and~$j=1, \dots, N'$, and~$N'' = N N'$.  The composite system is in a state represented by~$(p''_{l}, \var''_{l})$, where~$l=N'(i-1) + j$, and we assume that
\begin{subequations}
\begin{align}
p''_{l} &= p_i p'_j   		\label{eqn:composite-systems-1} \\
\var''_{l} &= g(\var_i, \var'_j),  \label{eqn:composite-systems-2}
\end{align}
\end{subequations}
where~$g$ is a function, symmetric in its arguments, to be determined.

Suppose that the first subsystem undergoes the gauge transformation~$(p_i; \var_i) \rightarrow (p_i; \var_i + \var_0)$.  We require that this transformation leads to a gauge transformation of the composite system, so that
\begin{equation} 
(p''_{l}, \var''_{l})  \rightarrow (p''_{l}, \var''_{l} + h(\var_0)),
\end{equation} 
where~$h$ is some function to be determined.  Together with Eq.~\eqref{eqn:composite-systems-1}, this implies that~$g$ is linear in its first argument.  Applying the same argument to the second subsystem, one obtains that~$g(\var_i, \var'_j) = c\var_i + d\var'_j + e$.   Imposing symmetry, and setting~$e=0$ without loss of generality, we obtain~$g(\var_i, \var'_j) = c(\var_i + \var'_j)$.  

To determine~$c$, we apply the Temporal Evolution postulate.   We require that, if the energies of the subsystems are~$E$ and~$E'$, respectively, the energy of the composite system is~$E+E'$.  From the Temporal Evolution postulate, it follows at once that~$c=1$.  Hence, we obtain~$g(\var_i, \var'_j) = \var_i + \var_j$.  Therefore, the state of the composite system~$(p''_{l}, \var''_{l}) = (p_i p_j'; \var_i + \var'_j)$, which is the content of the Composite Systems postulate.  

Alternatively, one can obtain this result more directly from the Hamilton-Jacobi model.  We note that, if, with respect to position measurements along the~$x$ and~$y$ axes, the discretized Hamilton-Jacobi state
of a particle is~$(p_i; S_i)$ and~$(p'_j; S'_j)$, respectively, where~$i=1, \dots, N$ and~$j=1, \dots, N'$, then, with respect to $(x,y)$-position measurements, its state is~$(p''_{l}; S''_l) =
(p_i p'_j; S_i + S'_j)$ where~$l= N(i-1)+j$.  By the classical-quantum correspondence, and then generalizing to the abstract quantum model,~$\model{q}(N)$, we obtain the same result.

\subsection{Novel Postulates}

\subsubsection*{Postulate~2.1: Complementarity}

According to the discussion of correspondence above, the state~$S(t)$,
written with respect to some measurement~$\mment{A} \in \set{A}$,
can be represented by the pair~$(\vect{p}, \vect{\var})$, where~$\vect{p}$
contains the probabilities of the measurement results,
and~$\vect{\var}$ is an ordered set of real-valued degrees of
freedom.  Hence, the state consists of a mixture of probabilities
and degrees of freedom unconnected to probabilities, and measurement~$\mment{A}$ yields information about the~$p_i$ but not the~$\var_i$. The Complementarity
postulate is motivated by the aesthetic desideratum the the quantum state, as far as possible, should consist of probabilities of events rather than being such a mixture, and aims to express the restriction on measurement~$\mment{A}$ as a restriction on the ability of measurement~$\mment{A}$ to completely resolve the events that occur when it is performed. 

In particular, we hypothesize that, when measurement~$\mment{A}$ is performed, there are, in 
fact, $2N$~possible outcomes, with respective probabilities~$P_1, \dots, P_{2N}$, and that result~$i$ is observed whenever either outcome~$2i-1$ or outcome~$2i$ is realized.  We note that similar assumptions have been made in toy models of quantum theory in order to give concrete expression to complementarity~\cite{Spekkens-toy-model}.

In Sec.~\ref{sec:discussion-of-postulate-2-2}, we sketch some
ideas which help to provide a better physical understanding of
this postulate.

\subsubsection*{Postulate~2.2: States}

The States postulate asserts that the state of a system with respect to measurement~$\mment{A} \in \set{A}$ is given by~$\lvect{Q} = (Q_1, \dots, Q_{2N})\transpose$, with~$Q_q \in [-1,1]$, where~$P_q = Q_q^2$.   Hence, in addition to the~$P_q$, this postulate asserts that there is an additional, binary degree of freedom,~$\sigma_q$, associated with each outcome~$q$, where~$\sigma_q = \operatorname{sign}(Q_q)$, which is defined whenever~$P_q \neq 0$. 

The motivation for the introduction of the~$\sigma_q$ is the following.  If one takes the~$\lvect{P}$ themselves as the state space of the system, one finds that non-trivial one-to-one transformations of the state space that preserve the metric over the state space~(as we shall require in the Metric Preservation postulate, described below) are not possible.  A simple way to allow the existence of such transformations is to take all~$\lvect{Q}$ as the state space of the system, for then there exist transformations, such as orthogonal transformations of the~$\lvect{Q}$, which preserve the metric over the~$\lvect{Q}$.

\subsubsection*{Postulate~2.3: State Representation}

The State Representation postulate connects together the hypothesis, expressed by the States postulate, that the state of a system is given by~$\lvect{Q} = (Q_1, \dots, Q_{2N})$ and the above assertion that the state can be represented by~$(\vect{p}, \vect{\var})$.  Specifically,  labeling the~$2N$ possible outcomes of measurement~$\mment{A}$ as~$1a, 1b, \dots, Na, Nb$, we can redraw the probability tree of Fig.~\ref{fig:prob-tree-complementarity} as shown in Fig.~\ref{fig:prob-tree}, where the results~$1, 2, \dots, N$ of the measurement are shown in the upper level of the tree.  

If result~$i$ is observed, the experimenter does not know whether outcome~$ia$ or~$ib$ was realized, or what were their polarities.  The probability that~$ia$ was realized given that result~$i$ was obtained is denoted~$p_{a|i}$, and similarly the probability that~$ib$ was realized given result~$i$ is denoted~$p_{b|i}$.     We can encode these probabilities and polarities into the quantities~$q_{a|i}$ and~$q_{b|i}$ by requiring that~$p_{a|i} = q_{a|i}^2$ and~$p_{b|i} = q_{b|i}^2$, and that the polarities~$\sigma_{2i-1} = \operatorname{sign}(q_{a|i})$ and~$\sigma_{2i} = \operatorname{sign}(q_{b|i})$, the polarities being defined whenever the corresponding probabilities are non-zero.

The State Representation postulate now connects the~$q_{a|i}$ and~$q_{b|i}$ together with the~$\var_i$ by asserting that
\begin{equation}
\begin{aligned}
q_{a|i} &= f(\var_i) \\
q_{b|i} &= \tilde{f}(\var_i),
\end{aligned}
\end{equation} 
where~$f$ is not a constant function, and where the functions~$f$ and~$\tilde{f}$ are function, assumed differentiable, to be determined.

\subsubsection*{Postulate~2.4: Information Metric}

The Information Metric postulate asserts that the metric over the space of probability distributions~$\lvect{P}$ is the information metric which, as discussed in the Introduction, naturally arises if one considers questions of information gain.

\subsubsection*{Postulate~3.5: Metric Preservation}

The Metric Preservation postulate accords the metric over state space a fundamental place in the theoretical framework:~any transformation of state space is required to preserve the distance between any pair of nearby states.  This idea is similar to the premise of  Wigner's theorem, namely that transformations of the space of pure states of a quantum system preserve the Hilbert space angle between any two pure states, which can be interpreted as requiring that transformations preserve the distinguishability of any pair of states~\cite{Wootters-statistical-distance}.


\section{Deduction of the quantum formalism}
\label{sec:D}

The derivation will proceed as follows.
First, in Sec.~\ref{sec:D1}, using the States and Information Metric postulates, we represent the state of a system,~$\state{S}(t)$, as a unit vector in a $2N$-dimensional real Euclidean space,~$Q^{2N}$, and use two of the Transformations postulates~(Mappings and One-to-One) to find that transformations over state space are orthogonal transformations.

Second, in Sec.~\ref{sec:D2}, the Measure Invariance postulate is used to determine the
form of the functions~$f$ and~$\tilde{f}$ that are introduced in the State Representation postulate.  We then apply the remaining Transformations
postulates, which lead to the conclusion that transformations representing physical transformations are, in fact, represented by a subset of the orthogonal transformations of
the unit hypersphere,~$\hypersphere$, in~$Q^{2N}$.  We then show that these
transformations can, equivalently, be represented by the set of
unitary and antiunitary transformations of a suitably-defined
$N$-dimensional complex vector space.  Finally, we show that physical transformations
parameterized by continuous parameters are represented either by unitary or antiunitary transformations,
and that continuous transformations are represented by unitary transformations.

Third, in Sec.~\ref{sec:D4}, we draw upon the Measurement
Simulability postulate in order to obtain a representation of
measurements on a system, and, in Sec.~\ref{sec:D5}, use the Composite Systems
postulate to obtain the tensor product rule, which
determines the state of a composite system in terms of the states of
its subsystems.

Finally, in Sec.~\ref{sec:generalisations}, we generalize the formalism to allow
the description of measurements performed on subsystems of a composite system, and to allow the description of measurements with degenerate values.   Additionally, in Sec.~\ref{sec:Haar}, we obtain a metric over the space of pure states, and obtain a unitarily- and antiunitarily-invariant prior over~$\hypersphere$.

\subsection{States and Dynamics in $Q$-space}
\label{sec:D1}

\subsubsection{$Q$-space Representation of State Space}
\label{sec:Q-space}

According to the States postulate, the state of a system prepared using any measurement in the measurement set~$\set{A}$ can be represented by~$\lvect{Q} = (Q_1, Q_2, \dots, Q_{2N-1}, Q_{2N})$. 
Now, the Information Metric postulate assigns the metric
\begin{equation}
ds^2 = \frac{1}{4} \sum_{q=1}^{2N} \frac{dP_q^2}{P_q}
\end{equation}
over~$\lvect{P}$, which, from the relation~$P_q = Q_q^2$, implies that the metric over~$\lvect{Q}$ is Euclidean, namely
\begin{equation} \label{eqn:Q-space-metric}
ds^2 = \sum_q dQ_q^2.
\end{equation}

Hence, the state space of the system can be represented by the set of all unit vectors in a $2N$-dimensional real Euclidean space, which we will refer to as $Q$-space or~$Q^{2N}$.

\subsubsection{Representation of Physical Transformations}
\label{sec:orthogonals}

By the Mappings and One-to-One postulates, any physical transformation is represented by a one-to-one map,~$\map{M}$, over state space,~$Q^{2N}$.   Furthermore, by the Metric Preservation postulate,~$\map{M}$ must be an orthogonal transformation of the unit hypersphere,~$\hypersphere$, in~$Q^{2N}$.

\subsection{Correspondence, and Complex Form of States and Dynamics}
\label{sec:D2}

\subsubsection{Determination of functions~$f$ and~$\tilde{f}$}
\label{sec:form-of-f}

According to the State Representation postulate, the state~$\lvect{Q}$, of a system with respect to some measurement~$\mment{A}\in\set{A}$ can be written
\begin{equation} \label{eqn:Q-p-var-form}
\lvect{Q} = (\sqrt{p_1} q_{a|1}, \sqrt{p_1} q_{b|1}, \dots, \sqrt{p_N} q_{a|N}, \sqrt{p_N} q_{b|N}),
\end{equation}
where~$q_{a|i} = f(\var_i)$ and~$q_{b|i} = \tilde{f}(\var_i)$.  In order to determine the unknown functions~$f$ and~$\tilde{f}$, we first determine the metric over~$\lvect{Q}$ in 
terms of the~$p_i$ and~$\var_i$.  

Using Eq.~\eqref{eqn:Q-space-metric}, 
\begin{align}
ds^2  &= dQ_1^2 + dQ_2^2 + \dots + dQ_{2N-1}^2 + dQ_{2N}^2 \\
	&=  \sum_{i=1}^N \frac{1}{4}\frac{dp_i^2}{p_i} + p_i \left( f'^2(\var_i) + \tilde{f}'^2(\var_i) \right) d\var_i^2  \\
	&=  \sum_{i=1}^N \frac{1}{4}\frac{dp_i^2}{p_i} +  p_i \left[ \frac{ f'^2(\var_i)}{1-f^2(\var_i)} \right] 
		d\var_i^2,
\end{align}
where we have used the relation~$f^2(\var_i) + \tilde{f}^2(\var_i) = 1$ in the third line.  Defining the function~$F(\var_i) = f^2(\var_i)$, we can write the above as
\begin{equation}
ds^2 	= \frac{1}{4} \sum_{i=1}^N \frac{dp_i^2}{p_i} +  p_i \left[ \frac{F'^2(\var_i)}{F(\var_i)(1-F(\var_i))}  \right] d\var_i^2.
\end{equation}

The measure over~$(p_1, \dots, p_N; \var_1, \dots, \var_N)$ induced by this metric is proportional to the square-root of the determinant of the metric, and so is given by
\begin{equation}
\mu(\vect{p}, \vect{\var}) = \mu_0 \, \prod_{i=1}^N   \frac{F'(\var_i) }{\sqrt{F(\var_i)(1-F(\var_i))}} \, ,
\end{equation}
where~$\mu_0$ is a constant, which marginalizes to give
\begin{equation} \label{eqn:marginalized-measure}
\begin{aligned}
\mu_i(\var_i) &\equiv \idotsint \mu(\vect{p}, \vect{\var}) \, dV_{\bar{\imath}} \\
		& = c\mu_0  \frac{F'(\var_i) }{\sqrt{F(\var_i)(1-F(\var_i))}},
\end{aligned}
\end{equation}
with~$dV_{\bar{\imath}} \equiv dp_1 \dots dp_N \, d\var_1 \dots d\var_{i-1} d\var_{i+1} \dots d\var_N$,
as the measure over~$\var_i$, where~$c$ is given by
\begin{multline}
c = \left( \idotsint  dp_1\dots dp_N \right)  \\
	\times \left( \int   \frac{F'(\var) }{\sqrt{F(\var)(1-F(\var))}} \, d\var \right)^{N-1}.
\end{multline}

Now, using the Measure Invariance postulate,~$\mu_i(\var_i + \var_0)$ is given by
\begin{equation}
\begin{aligned}
&\quad\idotsint \mu(\vect{p}, \var_1, \dots, \var_{i-1}, \var_i + \var_0, \var_{i+1}, \dots, \var_N) \, dV_{\bar{\imath}} \\
			&= \idotsint \mu(\vect{p}, \tilde{\var}_1 + \var_0, \dots, \tilde{\var}_{i-1} +\var_0, \var_i + \var_0, \\
				&\quad\quad\quad\quad\quad\quad\quad\quad\quad\quad\quad\quad\tilde{\var}_{i+1} + \var_0, \dots, \tilde{\var}_N + \var_0) d\tilde{V}_{\bar{\imath}} \\
			&= \idotsint \mu(\vect{p}, \tilde{\var}_1, \dots, \tilde{\var}_{i-1}, \var_i, \tilde{\var}_{i+1}, \dots, \tilde{\var}_N) d\tilde{V}_{\bar{\imath}} \\
			&= \mu_i(\var_i),
\end{aligned}
\end{equation}
with~$d\tilde{V}_{\bar{\imath}} \equiv dp_1 \dots dp_N \, d\tilde{\var}_1 \dots d\tilde{\var}_{i-1} d\tilde{\var}_{i+1} \dots d\tilde{\var}_N$, where the variable substitution~$\tilde{\var}_j = \var_j - \var_0$ for~$j \neq i$ has been used to obtain the second line, and Eq.~\eqref{eqn:measure-constraint} has been used to obtain the third line.    Hence, the measure~$\mu_i(\var_i)$ is independent of~$\var_i$.   Therefore,
\begin{equation}
\frac{1}{\sqrt{F(\var)\left(1-F(\var)\right)}} \frac{dF(\var)}{d\var} = 2a,
\end{equation}
where~$a$ is a constant, which has the general solution
\begin{equation} \label{eqn:F-function}
F(\var) = \cos^2(a\var + b),
\end{equation}
where~$b$ is a constant.  The constant~$a$ is non-zero since, by the State Representation postulate, the function~$f(\var)$ is not a constant.  Hence, the functions~$f(\var)$ and~$\tilde{f}(\var)$ have the form
\begin{equation} \label{eqn:f-function}
\begin{aligned}
 f(\var) &= \pm \cos(a\var +b)  \\
 \tilde{f}(\var) &= \pm \sin(a\var +b), \\
\end{aligned}
\end{equation}
where the signs of~$f$ and~$\tilde{f}$ are undetermined.

We note that the freedom in the choice of signs of~$f$ and~$\tilde{f}$ can be absorbed into the 
choice of~$a$ and~$b$. That is, if one chooses the signs of~$f$ and~$\tilde{f}$ not to be both
positive, then this is equivalent to choosing positive signs
for~$f$ and~$\tilde{f}$ but changing the values of~$a$ and~$b$ to
some other values,~$a'$ and~$b'$, respectively. Specifically, if
one chooses the signs~$(+,-)$, then~$a' = -a$ and~$b' = -b+\pi$;
if~$(-,+)$, then~$a' = -a$ and~$b' = -b$; and, if~$(-,-)$,
then~$a' = a$ and~$b' = b+\pi$.   Therefore, without loss of generality, 
the signs can be both taken to be positive.

Defining~$\phi_i = a\var_i + b$, we can write~$q_{a|i} = \cos\phi_i$
and~$q_{b|i} = \sin\phi_i$, and therefore, from Eq.~\eqref{eqn:Q-p-var-form},
write the state of a
system with respect to some measurement~$\mment{A} \in \set{A}$ as
\begin{equation} \label{eqn:defn-of-lvect-Q-2}
\lvect{Q}   = (\sqrt{p_1} \cos\phi_1, \sqrt{p_1} \sin\phi_1, \ldots,
             \sqrt{p_N} \sin\phi_N)\transpose.
\end{equation}

\subsubsection{Mappings}
\label{sec:mappings}

In this section, the general form of mappings that represent
physical transformations of a system  will be determined. The
derivation will proceed in three steps:
\begin{enumerate}
    \item[(1)] Show that the imposition of the Gauge Invariance postulate
    restricts~$\map{M}$
    to a subset of the set of orthogonal transformations, and that these
    transformations can be recast as unitary or antiunitary transformations
    acting on a suitably-defined complex vector space.

    \item[(2)] Show that any unitary or antiunitary transformation represents
    an orthogonal transformation satisfying the One-to-One,
    Gauge Invariance, and Metric Preservation postulates.

    \item[(3)] Show using the Continuity postulate that a physical
    transformation which depends continuously
    upon a real-valued parameter n-tuple can be represented by either
    unitary or antiunitary transformations, and show using the Continuous
    Transformations postulate that a
    continuous physical transformation can only be represented by
    unitary transformations.
\end{enumerate}

\paragraph{Step 1: Imposition of the Gauge Invariance postulate.}
\label{sec:M-is-equivalent-to-unitary}

From Eq.~\eqref{eqn:defn-of-lvect-Q-2}, the Gauge Invariance postulate, and the relation~$\phi_i = a\var_i + b$ given above, it follows that the
probabilities~$p_1', p_2', \dots, p_N'$ of the results of measurement~$\mment{A}$
performed on a system in state~$\lvect{Q}' = \map{M}(\lvect{Q})$ are
unaffected if, in any state~$\lvect{Q} = (\sqrt{p_1}\cos\phi_1, \sqrt{p_1}\sin\phi_1, \dots, \sqrt{p_N}\cos\phi_N, \sqrt{p_N}\sin\phi_N)$ written down with respect to
measurement~$\mment{A}$, an arbitrary real constant,~$\phi_0$, is
added to each of the~$\phi_i$.   Our goal in this section is to determine the constraint imposed on~$\map{M}$ by this condition.

Since~$\map{M}$ is an orthogonal transformation~(Sec.~\ref{sec:orthogonals}), it can be
represented by the~$2N$--dimensional orthogonal matrix,~$M$. Under
its action, the vector~$\lvect{Q}$ transforms as
\begin{equation} \label{eqn:Q-dash-is-Q-rotated}
    \lvect{Q}' = \rmatrix{M} \lvect{Q}.
\end{equation}
In order to determine the most general permissible form of~$\rmatrix{M}$, it is suffices to consider two types of special case.

First consider the case where
all but one of the~$p_i$ are zero.  For concreteness, suppose that~$p_1 = 1$ and~$p_2, \dots, p_N$ are all zero.  In that case, from Eq.~\eqref{eqn:Q-dash-is-Q-rotated}, using the relation~$p_1' = Q_1'^2 + Q_2'^2$, we obtain
\begin{equation} \label{eqn:basic-P-k-dash1}
    \begin{split}
 p_1'   &= \frac{1}{2} \left[ (M_{11}^2 + M_{21}^2) + (M_{12}^2 + M_{22}^2)  \right] \\
        &\quad+
            \frac{1}{2} \left[(M_{11}^2 + M_{21}^2) - (M_{12}^2 + M_{22}^2)   \right]
             \cos 2\phi_1 \\
           &\quad+
            \left(M_{11}M_{12} + M_{21} M_{22}\right) \sin 2\phi_1.
    \end{split}
 \end{equation}
We require that~$p_1'$ remains unchanged as a result of the
addition of any constant~$\phi_0 \in \numberfield{R}$ to
the~$\phi_i$. However, a linear combination of the
functions~$\cos 2\phi_1$
and~$\sin 2\phi_1$ in which at least one of the
coefficients is non-zero is zero only on a discrete set of points.
Therefore, the coefficients of the
functions~$\cos 2\phi_1$
and~$\sin 2\phi_1$ must vanish, so that the
conditions
\begin{gather}\label{eqn:rotation-matrix-constraintsA}
M_{11}^2 + M_{21}^2 =  M_{12}^2 + M_{22}^2  \\
M_{11}M_{12} + M_{21} M_{22} =0
\end{gather}
must hold, which implies that
\begin{equation} 
\begin{pmatrix}
M_{11} & M_{12} \\
M_{21} & M_{22} 
\end{pmatrix}  =  \alpha_{11} 
	\begin{pmatrix}
	 \cos \varphi_{11}  & -\sigma_{11}\sin\varphi_{11} \\
          \sin \varphi_{11}  &  \sigma_{11}\cos\varphi_{11},\
        \end{pmatrix},
\end{equation}
where~$\sigma_{11} = \pm 1$, 
which is matrix composed of a enlargement matrix~(scale
factor~$\alpha_{11}$) and a rotation matrix if~$\sigma_{11}
=1$ or a reflection-rotation matrix~(that is, a matrix
representing a reflection followed by rotation) if~$\sigma_{11}
=-1$, with rotation angle~$\varphi_{11}$ in either case.

More generally, in the case where~$p_i =1$, one obtains
\begin{equation} \label{eqn:basic-P-k-dash_general}
    \begin{split}
 p_k'   &= \frac{1}{2} \left(\alpha_{ki} + \beta_{ki} \right) 
        +
            \frac{1}{2} \left(\alpha_{ki}-\beta_{ki} \right)
             \cos 2\phi_i \\
           &\quad+ \gamma_{ki} \sin 2\phi_i
    \end{split}
 \end{equation}
for~$k=1, \dots, N$, where
    \begin{equation}
    \begin{aligned} \label{eqn:matrix-definitionsA}
        \alpha_{ki} &= M_{2k-1, 2i-1}^2 + M_{2k, 2i-1}^2                           \\
        \beta_{ki} &= M_{2k-1, 2i}^2 + M_{2k,2i}^2                                 \\
        \gamma_{ki} &= M_{2k-1, 2i-1}M_{2k-1,2i} + M_{2k, 2i-1}M_{2k, 2i}.
    \end{aligned}
    \end{equation}
The invariance condition implies the conditions
\begin{equation}\label{eqn:rotation-matrix-constraintsA_general}
\alpha_{ki} = \beta_{ki} \quad \text{and} \quad \gamma_{ki} = 0
\quad \text{for all~$i,k$},
\end{equation}
which implies that~$\rmatrix{M}$ takes the form of an~$N$-by-$N$ array of two-by-two sub-matrices,
\begin{equation} \rmatrix{M} =
\begin{pmatrix} \label{eqn:restricted-rotation}
     T^{(11)} &  T^{(12)} & \dots &
        T^{(1N)} \\
     T^{(21)} &  T^{(22)} & \dots &
         T^{(2N)} \\
    \hdotsfor[2]{4}                 \\
     T^{(N1)} &  T^{(N2)} & \dots &
         T^{(NN)}
\end{pmatrix},
\end{equation}
where
\[ T^{(ij)} = \alpha_{ij}
        \begin{pmatrix}
          \cos \varphi_{ij}  & -\sigma_{ij}\sin\varphi_{ij} \\
          \sin \varphi_{ij}  &  \sigma_{ij}\cos\varphi_{ij}
        \end{pmatrix}
\]
is a two-by-two matrix composed of a enlargement matrix~(scale
factor~$\alpha_{ij}$) and a rotation matrix if~$\sigma_{ij}
=1$ or a reflection-rotation matrix if~$\sigma_{ij}
=-1$, with rotation angle~$\varphi_{ij}$ in either case.

Before considering the second type of special case, we note that, due to the special form of~$\rmatrix{M}$ above, it is possible to rewrite Eq.~\eqref{eqn:Q-dash-is-Q-rotated} in a simpler way.  The state~$\lvect{Q}$ can be faithfully represented by 
\begin{equation}  \label{eqn:complex-form-of-state}
    \cvect{v}=
    \begin{pmatrix}
        Q_1 + iQ_2 \\
        Q_3 + iQ_4 \\
        \dots \\
        Q_{2N-1}  + iQ_{2N}
    \end{pmatrix},
\end{equation}
so that, from Eq.~\eqref{eqn:defn-of-lvect-Q-2},~$v_i = \sqrt{p_i}\, e^{i\phi_i}$.  Define the complex matrix
\begin{equation}
\cmatrix{W} = \begin{pmatrix}
			\alpha_{11} e^{i\varphi_{11}} K^{\sigma_{11}} &\dots & \alpha_{1N} e^{i\varphi_{1N}} K^{\sigma_{1N}}  \\
			\alpha_{21} e^{i\varphi_{21}} K^{\sigma_{21}} &\dots & \alpha_{2N} e^{i\varphi_{2N}} K^{\sigma_{2N}}  \\
			\dots & \dots & \dots \\
			\alpha_{N1} e^{i\varphi_{N1}} K^{\sigma_{N1}} &\dots & \alpha_{NN} e^{i\varphi_{NN}} K^{\sigma_{NN}}
			\end{pmatrix},
\end{equation}
where~$K^\sigma$ is the conditional complex conjugation operation defined as
\begin{equation}
K^\sigma z = \begin{cases}
			z & \text{if~$\sigma =1$} \\
			z^* & \text{if~$\sigma =-1$},
			\end{cases}
\end{equation}
where~$z \in \numberfield{C}$.  Then, Eq.~\eqref{eqn:Q-dash-is-Q-rotated} is equivalent to the equation
\begin{equation}  \label{eqn:complex-form-of-Q-dash}
\cvect{v}' = \cmatrix{W} \cvect{v},
\end{equation}
where~$\cvect{v}'$ is defined analogously to~$\cvect{v}$.

Consider now the second special case, where two of the~$p_i$,
say~$p_i$ and~$p_j$~$(i\neq j)$ are set equal to~$1/2$, and the
remainder are set to zero.  For example, if~$p_1 = p_2 =1/2$, then Eq.~\eqref{eqn:complex-form-of-Q-dash} yields
\begin{equation} 
    \begin{split}
 p_1'   &= \frac{1}{4} \left| \alpha_{11} e^{i(\varphi_{11}  + \sigma_{11}\phi_1)} + \alpha_{12} e^{i(\varphi_{12}  + \sigma_{12}\phi_2)} \right|^2  \\
 	&= \frac{1}{4} \bigl\{ \alpha_{11}^2 + \alpha_{12}^2 \\
	&\quad\quad +2\alpha_{11}\alpha_{12} \cos\left[ (\varphi_{11} - \varphi_{12}) + (\sigma_{11} \phi_1 - \sigma_{12}\phi_2) \right] \bigr\}.
	\end{split}
\end{equation}
In order that~$p_1'$ remains unchanged as a result of
the addition of~$\phi_0 \in \numberfield{R}$ to the~$\phi_i$, either~$\alpha_{11}\alpha_{12} =0$ or~$\sigma_{11} = \sigma_{12}$.  

More generally, in the case where~$p_i = p_j =1/2$, with~$i \neq j$, one obtains the expression
\begin{equation}
\begin{split}
p_k' &= \frac{1}{4}  \bigl\{ \alpha_{ki}^2 + \alpha_{kj}^2 \\
	& - 2\alpha_{ki}\alpha_{kj} \cos\left[ (\varphi_{ki} - \varphi_{kj}) + (\sigma_{ki} \phi_i - \sigma_{kj}\phi_j) \right] \bigr\},
\end{split}
\end{equation}
and the invariance condition implies that, for any~$k$ and any~$i\neq j$, the values of~$\sigma_{ki}$ and~$\sigma_{kj}$ must be the same unless~$\alpha_{ki}=0$ or~$\alpha_{kj}=0$.

Since~$\rmatrix{M}$ represents the mapping~$\map{M}$, and, by the
One-to-One postulate,~$\map{M}^{-1}$ exists, the
matrix~$\rmatrix{M}^{-1}$ represents the mapping~$\map{M}^{-1}$.
Hence, the matrix~$\rmatrix{M}^{-1} = \rmatrix{M}^\textsf{T}$ must also
satisfy the Invariance postulate.  Now, from
Eq.~\eqref{eqn:restricted-rotation}, the matrix~$\rmatrix{M}^\textsf{T}$
takes the form
\begin{equation}
\rmatrix{M}^\textsf{T} =
                \begin{pmatrix}
                \label{eqn: restricted-rotation}
                \left(T^{(11)}\right)^\textsf{T} &  \left(T^{(21)}\right)^\textsf{T} & \dots &
                \left(T^{(N1)}\right)^\textsf{T} \\
                \left(T^{(12)}\right)^\textsf{T} &  \left(T^{(22)}\right)^\textsf{T} & \dots &
                \left(T^{(N2)}\right)^\textsf{T} \\
                    \hdotsfor[2]{4}                 \\
                 \left(T^{(1N)}\right)^\textsf{T} &  \left(T^{(2N)}\right)^\textsf{T} & \dots &
                \left(T^{(NN)}\right)^\textsf{T}
\end{pmatrix},
\end{equation}
and the corresponding complex matrix is
\begin{equation}
\widetilde{W}_{ij} = \alpha_{ji} \, e^{-i\sigma_{ji}\varphi_{ji}} K^{\sigma_{ji}}.
\end{equation}

Consider the transformation~$\cvect{v}' = \widetilde{\cmatrix{W}} \cvect{v}$, with the above special case, namely~$p_i = p_j = 1/2$, where~$i\neq j$.    In this case, one obtains
\begin{equation}
\begin{split}
p_k' &= \frac{1}{4}  \bigl\{ \alpha_{ik}^2 + \alpha_{jk}^2 \\
& - 2\alpha_{ik}\alpha_{jk} \cos\left[ (-\sigma_{ik}\varphi_{ik} + \sigma_{jk} \varphi_{jk}) + (\sigma_{ik} \phi_i - \sigma_{jk}\phi_j) \right] \bigr\}.
\end{split}
\end{equation}
The invariance condition implies that, for any~$k$ and any~$i\neq j$, the values of~$\sigma_{ik}$ and~$\sigma_{jk}$ must be the same unless~$\alpha_{ik}=0$ or~$\alpha_{jk}=0$.  But, this implies that, in~$\cmatrix{W}$, all of the non-zero entries have the same value of~$\sigma_{ik}$.  Therefore, the form of~$\cmatrix{W}$ is of one of two types,
\begin{equation} \label{eqn:W-types}
\cmatrix{W} =  \cmatrix{V}
			\quad\text{or}\quad
\cmatrix{W} = \cmatrix{V}\cmatrix{K},
\end{equation}
corresponding to the cases where the~$\sigma_{ij} = 1$ and the~$\sigma_{ij}=-1$, respectively, where
\begin{equation}
{V}_{ij} = \alpha_{ij} e^{i\varphi_{ij}},
\end{equation}
and~$\cmatrix{K}$ is the complex conjugation operator,~$\cmatrix{K}\cvect{v} = \cvect{v}^*$.

Now, since~$\rmatrix{M}$ is orthogonal, it follows that~$\cmatrix{V}$ is unitary. To see this, consider~$\rmatrix{M}$ with all~$\sigma_{ij} =1$.  In that case, 
\begin{equation}  \label{eqn:orthogonality_of_M}
\begin{split}
  (\rmatrix{M}\transpose \rmatrix{M})_{[ij]} &= \sum_k \left( \rmatrix{T}^{(ki)} \right)^{\text{\textsf{T}}}  \rmatrix{T}^{(kj)} \\
  				&= \sum_k \alpha_{ki} \alpha_{kj}  \rmatrix{R}(-\varphi_{ki})  \rmatrix{R}(\varphi_{kj}) \\
				&=  \sum_k \alpha_{ki} \alpha_{kj}  \rmatrix{R}(\varphi_{kj} - \varphi_{ki})
\end{split}
\end{equation}
where~$\rmatrix{A}_{[ij]}$ denotes the~$(i,j)$th two-by-two submatrix of~$\rmatrix{A}$, with~$\rmatrix{A}$ being an $N$~by~$N$ array of two-by-two submatrices, and where~$\rmatrix{R}(\varphi)$ is a two-by-two rotation matrix with rotation angle~$\varphi$.  Consider also
\begin{equation} \label{eqn:VVij}
    (\cmatrix{V}^\dag \cmatrix{V})_{ij} =
            \sum_k \alpha_{ki}  \alpha_{kj}e^{i(\varphi_{kj} - \varphi_{ki})}.
\end{equation}
By inspection, the orthogonality condition~$(\rmatrix{M}\transpose \rmatrix{M})_{[ij]} = \delta_{ij} \rmatrix{I}$, where~$\rmatrix{I}$ is the two-by-two identity matrix, is equivalent to the condition of unitarity,~$(\cmatrix{V}^\dag \cmatrix{V})_{ij} = \delta_{ij}$.   Therefore, since~$\rmatrix{M}$ is orthogonal,~$\cmatrix{V}$ is unitary.  Hence, from Eq.~\eqref{eqn:W-types}, matrix~$\cmatrix{W}$ is either unitary or antiunitary.

Finally, we must show that transformations of the types in Eq.~\eqref{eqn:W-types} satisfy the
Gauge Invariance postulate for any state~$\lvect{Q}$, not just for the special cases of~$\lvect{Q}$ considered above.  This follows immediately:~we note that the addition
of~$\phi_0$ to each of the~$\phi_i$ in the complex form of the
state,~$\cvect{v}$, generates the state~$e^{i\phi_0}\, \cvect{v}$,
that is
\begin{equation} \label{eqn:invariance-of-P'-A}
\cvect{v} \xrightarrow{+ \phi_0} e^{i\phi_0} \,\cvect{v}.
\end{equation}
As a result, the vector~$\cvect{v}' = \cmatrix{V}\cvect{v}$
transforms as
\begin{equation} \label{eqn:invariance-of-P'-B}
\cvect{v}' \xrightarrow{+ \phi_0} e^{i\phi_0} \,\cvect{v}',
\end{equation}
and the vector~$\cvect{v}' = \cmatrix{V}\cmatrix{K}\cvect{v}$ transforms as
\begin{equation} \label{eqn:invariance-of-P'-C}
\cvect{v}' \xrightarrow{+ \phi_0} e^{-i\phi_0} \,\cvect{v}',
\end{equation}
Since~$p_i' = |v_i'|^2$, the~$p_i'$ are independent of the overall phase
of~$\cvect{v}'$, so that, in both
Eqs.~\eqref{eqn:invariance-of-P'-B}
and~\eqref{eqn:invariance-of-P'-C}, the~$p_i'$ remain unchanged by
the addition of~$\phi_0$ to the~$\phi_i$. Therefore, the
transformations~$\cmatrix{V}$ and~$\cmatrix{V}\cmatrix{K}$ both
satisfy the Gauge Invariance postulate for all~$\lvect{Q}$.

\paragraph{Step 2: General Unitary and Antiunitary Transformations.}
\label{sec:every-unitary-is-ok}

We have shown so far that the imposition of the Gauge Invariance postulate
restricts~$\rmatrix{M}$ to a subset of the set of orthogonal
transformations, and that each transformation in this subset can be
recast either as a unitary transformation or 
as an antiunitary transformation.
However, we have not ruled out the possibility that there exist
unitary or antiunitary transformations which are not equivalent to
orthogonal transformations belonging to the above mentioned subset. In
this section, it shall be shown that, in fact, \emph{any}
$N$-dimensional unitary or antiunitary transformation satisfies the
One-to-One, Metric Preservation, and Gauge Invariance postulates.

Consider first the arbitrary unitary transformation~$\cmatrix{U}$, where~$U_{ij} = \alpha_{ij} e^{i\varphi_{ij}}$.  The
transformation
\begin{equation} \label{eqn:general-U-on-v}
\cvect{v}' = \cmatrix{U} \cvect{v}
\end{equation}
is equivalent to the transformation
\begin{equation} \lvect{Q}' =
\rmatrix{M} \lvect{Q},
\end{equation}
where~$M_{[ij]} = \alpha_{ij} \rmatrix{R}(\varphi_{ij})$.
Now, as we observed above, the condition of unitarity of~$\cmatrix{U}$, namely~$(\cmatrix{U}^\dagger \cmatrix{U})_{ij} = \delta_{ij}$, is equivalent to the orthogonality condition of~$\rmatrix{M}$.  Therefore,~$\rmatrix{M}$ is orthogonal.

Similarly, in the case of the arbitrary antiunitary transformation~$\cmatrix{UK}$, where~$\cmatrix{U}$ is defined as above and~$\cmatrix{K}$ is the complex conjugation operator, the corresponding matrix~$\rmatrix{M}$ is given by~$M_{[ij]} = \alpha_{ij} \rmatrix{R}(\varphi_{ij}) \rmatrix{F}$, where~$\rmatrix{F}= \begin{pmatrix} 1 & 0 \\ 0 & -1 \end{pmatrix}$.  In this case, 
\begin{equation}
\begin{split}
\left( \rmatrix{M}\transpose \rmatrix{M} \right)_{[ij]}&= \sum_k \left(\rmatrix{M}\transpose \right)_{[ik]} \rmatrix{M}_{[kj]} \\
									&= \sum_k \alpha_{ki}\alpha_{kj} \rmatrix{F}\rmatrix{R}(-\varphi_{ki}) \rmatrix{R}(\varphi_{kj}) \rmatrix{F} \\
									&=\sum_k \alpha_{ki}\alpha_{kj} \rmatrix{R}(\varphi_{ki} - \varphi_{kj}),
\end{split}
\end{equation}
which is~$\delta_{ij} \rmatrix{I}$ due to the unitarity of~$\cmatrix{U}$.  Therefore,~$\rmatrix{M}$ is orthogonal in this case also.

Since~$\rmatrix{M}$ is an orthogonal matrix, it satisfies the
One-to-One and Metric Preservation postulates. The invariance of the~$p_i'$
required by the Gauge Invariance postulate follows from the observation made previously 
that, under the addition of~$\phi_0$ to the~$\phi_i$ in~$\cvect{v}$, the transformed vectors~$\cvect{v}' = \cmatrix{U}\cvect{v}$ or~$\cvect{v}' = \cmatrix{UK}\cvect{v}$ yield unchanged values of~$p_i'$.

Hence, any unitary or antiunitary transformation satisfies the
One-to-One,  Metric Preservation, and Gauge Invariance postulates.   In particular, we have obtained the result of Wigner's theorem that a one-to-one map over state space that represents a symmetry transformation of a system is either unitary or antiunitary.

\paragraph{Step 3: Physical Transformations.}

By the Continuity postulate, a physical transformation~(such as a
reflection-rotation of a frame of reference) that depends
continuously upon a real-valued parameter n-tuple~$\Bpi$ is
represented by a map~$\map{M}_{\Bpi}$ which depends continuously
upon~$\Bpi$.  From the above discussion, the set of mappings that represent physical transformations are the  set of all unitary and antiunitary transformations.  The set of all unitary transformations and the set of all antiunitary transformations are disjoint and it not possible to continuously transform a unitary transformation into any antiunitary transformation.  Therefore,~$\map{M}_{\Bpi}$ is represented either by unitary transformations or by antiunitary transformations.

Furthermore, by the Continuous Transformations postulate, a
\emph{continuous} physical transformation that depends continuously
upon a real-valued parameter n-tuple~$\Bpi$ is represented by a
map~$\map{M}_{\Bpi}$ which reduces to the identity map for some
value of~$\Bpi$.  However, only the set of unitary transformations include the identity.   Therefore, a continuous physical transformation can only be
represented by unitary transformations.  In particular, since 
temporal evolution of a system is a continuous physical transformation, it must be
represented by unitary transformations.

\subsection{Representation of Measurements} \label{sec:D4}

In the previous section, it has been shown that the state of a
system at time~$t$ that has been prepared by a measurement
in~$\set{A}$ can, from the point of view of a measurement~$\mment{A}
\in \set{A}$, be represented by the complex vector
\begin{equation}
\cvect{v} =(\sqrt{p_1} \,e^{i\phi_1}, \sqrt{p_2} \,e^{i\phi_2}, \dots, \sqrt{p_N} \,e^{i\phi_N})\transpose,
\end{equation}
where the~$p_i$ are the probabilities of
the results of measurement~$\mment{A}$ if performed at time~$t$. Furthermore, it
has been shown that any interaction following the preparation can
be represented by a unitary transformation of~$\cvect{v}$.

Consider an experiment where a system undergoes some
measurement~$\mment{A} \in \set{A}$, yields a particular result,
and subsequently undergoes some other measurement~$\mment{A}' \in
\set{A}$ that may or may not be the same as~$\mment{A}$.  The
purpose of this section is to develop the formalism necessary to
predict the measurement probabilities in such an experiment.

\subsubsection{Prepared States}

Suppose that, in the above-mentioned experiment, a system
undergoes measurement~$\mment{A}$ and yields result~$j$. What is
the state of the prepared system?

By the Results postulate, measurement~$\mment{A}$ has~$N$
possible results and, by the assumption of
reproducibility~(Sec.~\ref{sec:F-idealised-set-up}),
after~$\mment{A}$ has been performed and result~$j$ obtained,
immediate repetition yields the same result with certainty.
Therefore, for every result~$j$ there exists a corresponding
state,~$\cvect{v}_j$, such that the measurement~$\mment{A}$ upon the
system in state~$\cvect{v}_j$ yields result~$j$ with certainty.
From Eq.~\eqref{eqn:complex-form-of-state}, since~$p_j = 1$ and all
the other~$p_j$ are zero, we have that
\begin{equation} \label{eqn:form-of-v-j}
\cvect{v}_j = ( 0,\dots, e^{i\phi_j}, \dots, 0)^{\text{\textsf{T}}},
\end{equation}
where~$\phi_j$ is undetermined.

\subsubsection{Measurements}

\begin{figure}[!h]
\begin{centering}
\includegraphics[width=3.25in]{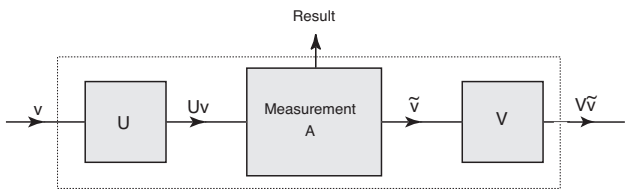}
\caption{\label{fig:representation-of-measurement} 
Simulation of measurement~$\mment{A}'$.  A unitary
transformation,~$\cmatrix{U}$, transforms the input
state,~$\cvect{v}$, into~$\cmatrix{U}\cvect{v}$.
Measurement~$\mment{A}$ is performed on this state, and yields a result and the outgoing
state,~$\tilde{\cvect{v}}$, which is then transformed by
the unitary transformation~$\cmatrix{V}$
into the output state~$\cmatrix{V}\tilde{\cvect{v}}$.}
\end{centering}
\end{figure}

By the Measurement Simulability postulate,
measurement~$\mment{A}'$ can be simulated by an arrangement
consisting of a measurement~$\mment{A}$ followed immediately before
and after by suitable interactions. These interactions bring about
continuous transformations of the system. From the results of the
previous section, these interactions must, therefore, be represented
by unitary transformations, which we shall denote~$\cmatrix{U}$
and~$\cmatrix{V}$, respectively~(see
Fig.~\ref{fig:representation-of-measurement}).  In the following, we
shall establish the form of these matrices, and then obtain an
expression for the probabilities for
measurement~$\mment{A}'$ performed on a system in state~$\cvect{v}$.

First, from the Results postulate and the assumption of
reproducibility, there exist~$N$ states~$\cvect{v}_1', \cvect{v}_2',
\ldots, \cvect{v}_N'$ such that measurement~$\mment{A}'$ performed
on a system in state~$\cvect{v}_i'$ yields result~$i$ with
certainty. Hence, the arrangement in
Fig.~\ref{fig:representation-of-measurement} must be such
that~$\mment{A}$ yields result~$i$ with certainty when the input
state to the arrangement is~$\cvect{v}_i'$.  For this to be the
case,~$\cmatrix{U}$ must transform~$\cvect{v}_i'$ to a state of the
form~$\cvect{v}_i e^{i\vard_i}$, where~$\vard_i$ is arbitrary. That
is, the matrix~$\cmatrix{U}$ must satisfy the relations
\begin{equation} \label{eqn:measurement-relations-for-U}
    \cmatrix{U} \cvect{v}_i' = \cvect{v}_i e^{i\vard_i},
    \quad i =1,2, \dots, N
\end{equation}

Second, if result~$i$ is obtained from the arrangement, the
output state of the arrangement must be of the form~$\cvect{v}_i'
e^{i\vard_i'}$, where~$\vard_i'$ is arbitrary.  But, immediately
after measurement~$\mment{A}$, the system is in
state~$\cvect{v}_i$ up to an overall phase.  Hence, the
matrix~$\cmatrix{V}$ must satisfy the relations
\begin{equation} \label{eqn:measurement-relations-for-V}
    \cmatrix{V} \cvect{v}_i = \cvect{v}_i' e^{i \vard_i'}
    \quad i =1,2, \dots, N
\end{equation}

From Eq.~\eqref{eqn:form-of-v-j}, the~$\cvect{v}_i$ form an
orthonormal basis for~$\numberfield{C}^N$, and, from
Eq.~\eqref{eqn:measurement-relations-for-U},~$\cvect{v}_i' =
\cmatrix{U}^\dag \cvect{v}_i e^{i\vard_i}$, which,
since~$\cmatrix{U}$ is unitary, implies that the~$\cvect{v}_i'$ also
form an orthonormal basis.  Therefore, any state,~$\cvect{v}$, can
be expanded as~$\sum_i c_i' \cvect{v}_i'$, with~$c_i' \in
\numberfield{C}$, and the matrices~$\cmatrix{U}$ and~$\cmatrix{V}$
are determined by the relations in
Eqs.~\eqref{eqn:measurement-relations-for-U}
and~\eqref{eqn:measurement-relations-for-V} up to the~$\vard_i$ and
the~$\vard_i'$.

It is now possible to determine the measurement probabilities if a
system in state~$\cvect{v}$ undergoes measurement~$\mment{A}'$.
Using Eq.~\eqref{eqn:measurement-relations-for-U} and the
expansion~$\cvect{v} = \sum_i c_i' \cvect{v}_i'$, the first
interaction of the arrangement transforms~$\cvect{v}$ into
\begin{equation}
\cmatrix{U}\left(\sum_i c_i'\cvect{v}_i'\right) = \sum_i
c_i'\cvect{v}_i e^{i\vard_i}.
\end{equation}
The probability that measurement~$\mment{A}$ in the arrangement
yields result~$i$ is therefore~$|c_i'|^2$. Hence,
measurement~$\mment{A}'$ performed on the state~$\cvect{v}$ yields
result~$i$ with probability~$|c_i'|^2$.  If result~$i$ is obtained, the outgoing state of measurement~$\mment{A}$ is~$\cvect{v}_i$, so that the output state of the arrangement is~$\cmatrix{V}\cvect{v}_i$ which is~$\cvect{v}'_i$ up to an overall phase.

In summary, every measurement,~$\mment{A}' \in \set{A}$, has an
associated orthonormal basis,~$\{ \cvect{v}_1', \cvect{v}_2',
\dots, \cvect{v}_N' \}$.  Such a measurement can be simulated by
a measurement~$\mment{A}$ followed immediately before and after by
interactions represented by~$\cmatrix{U}$ and~$\cmatrix{V}$
defined in Eqs.~\eqref{eqn:measurement-relations-for-U}
and~\eqref{eqn:measurement-relations-for-V} in terms of these
basis vectors. If measurement~$\mment{A}'$ is performed upon a
system in state~$\cvect{v}$, the probability,~$p_i'$, of obtaining
result~$i$ and corresponding output state~$\cvect{v}'_i$ is~$|c_i'|^2$, where~$c_i'$ is determined by the
relation~$\cvect{v} = \sum_i c_i' \cvect{v}_i'$, which is the Born rule.

\subsubsection{Expected Values}

If the $i$th~result of measurement~$\mment{A}'$ has an associated
real value~$a_i'$, the expected value obtained in an experiment in
which a system in state~$\cvect{v}$ undergoes
measurement~$\mment{A}'$ is defined as
\begin{equation}
\langle \cmatrix{A}' \rangle = \sum_i a_i' p_i'.
\end{equation}
Since~$p_i' = |c_i'|^2$ and~$c_i' = \cvect{v}_i'^\dag \cvect{v}$,
this expression can be also written as
\begin{equation}
\begin{split}
\langle \cmatrix{A}' \rangle &= \sum_i \cvect{v}^\dag
                    \left(\cvect{v}_i'
                        a_i'\cvect{v}_i'^\dag \right) \cvect{v} \\
                    &= \cvect{v}^\dag \left(\sum_i \cvect{v}_i'
                        a_i'\cvect{v}_i'^\dag \right) \cvect{v} \\
                    &= \cvect{v}^\dag \cmatrix{A}' \cvect{v},
\end{split}
\end{equation}
where the matrix~$\cmatrix{A}' \equiv \sum_i \cvect{v}_i'
a_i'\cvect{v}_i'^\dag$ is Hermitian since the~$a_i'$ are real, and
is non-degenerate since the~$a_i'$ have been assumed to be
distinct~(Sec.~\ref{sec:F-idealised-set-up}).

Since the~$\cvect{v}_i'$ are eigenvectors of~$\cmatrix{A}'$, with
the~$a_i'$ being the corresponding eigenvalues, the
matrix~$\cmatrix{A}'$ provides a compact mathematical way of
representing all the relevant details about
measurement~$\mment{A}'$.

\subsection{Composite Systems}
\label{sec:D5}

It is often the case that a given physical system
can be subject to examination in distinct experimental set-ups,
where the measurements in each set-up probe
distinct aspects of the system.  Formally, we can express this
as follows.

Consider a system which admits abstract quantum
model,~$\model{q}(N)$, with respect to measurement
set~$\set{A}^{(1)}$, and which admits abstract quantum
model,~$\model{q}(N')$, with respect to measurement
set~$\set{A}^{(2)}$, where the set-ups defined by measurement
sets~$\set{A}^{(1)}$ and~$\set{A}^{(2)}$ are disjoint~(in the sense
defined in Sec.~\ref{sec:F}). The system can also be modeled as a
whole. That is, we can construct the measurement set~$\set{A} =
\set{A}^{(1)} \times \set{A}^{(2)}$, and construct abstract quantum
model~$\model{q}(N'')$, where~$N'' = NN'$. We shall
accordingly speak of the system as a \emph{composite} system
consisting of two \emph{subsystems}. More generally, if a system
admits~$d$~($d>1$) abstract quantum models with respect to
$d$~disjoint measurement sets, we shall speak of it as a composite
system consisting of $d$~subsystems.

One often prepares a state of a composite system by first preparing
each of its subsystems, and then allowing these subsystems to
interact with one another.  In order to formally describe such a
procedure, one needs a rule, the composite system rule, which we
shall now derive, that enables the state of the system to be written
down in terms of the states of its subsystems.

\subsubsection*{The Composite System Rule}

In order to derive the composite system rule, we shall apply the
Composite Systems postulate to the case of a composite system with
two subsystems with abstract models~$\model{q}(N)$
and~$\model{q}(N')$, respectively, where the composite system
has the abstract model~$\model{q}(N'')$.

Suppose that the subsystems are in states represented
by~$(p_i; \var_i)$ and~$(p_j'; \var_j')$,
respectively, and the state of the composite system is represented by~$(p_l''; \var_l'')$~$(l=1, \dots, N'')$ in the manner defined by the Composite Systems postulate.
If we write the states of the subsystems in complex form,
        \begin{equation*}\cvect{v}^{(1)} = \left(
                             \sqrt{p_1}\,e^{i\phi_1},
                             \sqrt{p_2}\,e^{i\phi_2},
                             \dots,
                             \sqrt{p_N}\,e^{i\phi_N}\right)^{\text{\textsf{T}}}
        \end{equation*}
and
        \begin{equation*}
        \cvect{v}^{(2)} = \left(
                            \sqrt{p_1'}\,e^{i\phi_1'},
                            \sqrt{p_2'}\,e^{i\phi_2'},
                            \dots,
                            \sqrt{p_{N'}'} \, e^{i\phi_{N'}'}\right)^{\text{\textsf{T}}},
        \end{equation*}
respectively, where~$\phi_i = a\var_i + b$ and~$\phi'_j =a\var'_j+b$, and, similarly, write the state of the composite system as
        \begin{equation*}
        \cvect{v} = \left(
                             \sqrt{p_1''}\,e^{i\phi_1''},
                             \sqrt{p_2''}\,e^{i\phi_2''},
                             \dots,
                             \sqrt{p_{N''}''} \, e^{i\phi_{N''}''} 
                             \right)^{\text{\textsf{T}}},
        \end{equation*}
where~$\phi''_l=a\var''_l + b$, then it follows from the Composite Systems postulate that~$\cvect{v}$ can simply be
written as~$\cvect{v}^{(1)} \otimes \cvect{v}^{(2)}$.

More generally, consider a composite system with $d$~subsystems,
numbered~$1, 2, \dots, d$, in states~$\cvect{v}^{(1)},
\cvect{v}^{(2)}, \dots, \cvect{v}^{(d)}$, respectively.  We can
regard subsystems~$1$ and~$2$ as comprising a bipartite composite
system, system~$1'$, which, according to the above result, is in
state~$\cvect{v}^{(1)} \otimes \cvect{v}^{(2)}$. Next, we can
regard system~$1'$ and subsystem~$3$ as comprising a bipartite
composite system, system~$2'$, which is therefore in
state~$(\cvect{v}^{(1)} \otimes \cvect{v}^{(2)}) \otimes
\cvect{v}^{(3)}$.  Continuing in this way, we can see the state of
the composite system with $d$~subsystems has the state~$\cvect{v}
= \cvect{v}^{(1)} \otimes \cvect{v}^{(2)} \otimes \dots \otimes
\cvect{v}^{(d)}$.

\subsection{Some Generalizations} \label{sec:generalisations}

\subsubsection{Representation of subsystem measurements}

Suppose that measurement~$\mment{A}^{(1)} \in \set{A}^{(1)}$,
represented by $N$-dimensional Hermitian
operator~$\cmatrix{A}^{(1)}$, with eigenstates~$\cvect{v}_i^{(1)}$
and eigenvalues~$a_i$, respectively, is performed on subsystem~1 of dimension~$N$
of a bipartite composite system of dimension~$N''=NN'$, where~$N'$ is the dimension of 
subsystem~2. With respect to the abstract
quantum model~$\model{q}(N'')$ of the composite system,
measurement~$\mment{A}^{(1)}$ is \emph{not} in the measurement
set~$\set{A}$ of the composite system since the measurement has
only~$N$ distinct results whereas a measurement
in~$\set{A}$ has~$N'' = N N' > N$ possible
results.  However, it is convenient to be able to describe
measurement~$\mment{A}^{(1)}$, which we shall describe as a
\emph{subsystem measurement}, as an $N''$-dimensional
operator~$\cmatrix{A}$, in the framework of~$\model{q}(N'')$.

To determine the form of~$\cmatrix{A}$, it is sufficient to consider
the effect of~$\cmatrix{A}$ on product states of the
form~$\cvect{v}_i^{(1)} \otimes \cvect{v}^{(2)}$ of the composite
system, where~$\cmatrix{A}^{(1)}\cvect{v}_i^{(1)} = a_i
\cvect{v}_i^{(1)}$.  If the composite system is in such a state,
then subsystem~1 is in state~$\cvect{v}_i^{(1)}$. Therefore, when
measurement~$\mment{A}^{(1)}$ is performed, result~$a_i$ is
obtained with certainty, and the state of subsystem~1 is
unchanged~(up to an irrelevant overall phase). Therefore, the state
of the composite system remains unchanged.  If we require
that~$\cmatrix{A}$ has eigenvectors~$\cvect{v}_i^{(1)} \otimes
\cvect{v}^{(2)}$, with respective eigenvalues~$a_i$, it follows
that~$\cmatrix{A}$ can be taken to be~$\cmatrix{A}^{(1)} \otimes
\cmatrix{I}^{(2)}$, where~$\cmatrix{I}^{(2)}$ is the identity matrix
in the model of subsystem~2, with the only freedom being a
physically irrelevant overall phase in each of the eigenstates
of~$\cmatrix{A}^{(1)}$.

The above result trivially generalizes to the case of a measurement
performed on one subsystem of a composite system consisting of
$d$~subsystems.

\subsubsection{Degenerate measurements}
\label{sec:non-distinctness}

The model~$\model{q}(N)$, whose explicit mathematical form has been
derived above, applies to an abstract set-up where the measurements,
chosen from the set~$\set{A}$, have~$N$ possible results and
therefore, by the distinctness assumption of
Sec.~\ref{sec:F-idealised-set-up}, necessarily have~$N$ distinct
values. From the above discussions, it follows that each
measurement~$\mment{A} \in \set{A}$ is represented by a
non-degenerate Hermitian operator of dimension~$N$.

Now, it is useful to be able to describe measurements, within the
context of model~$\model{q}(N)$, which have fewer than~$N$ results.
An example of such measurements that we have discussed above are
subsystem measurements.  We shall now broaden the discussion to
allow for measurements with~$N'<N$ possible results where~$N'$ is
not a multiple of~$N$ and which therefore cannot be regarded as
subsystem measurements.

Consider an abstract set-up where a preparation implemented using a
measurement from~$\set{A}$ is followed by measurement~$\mment{A}$,
whose probabilities are denoted~$p_1, \dots,
p_N$. Suppose that, if measurement~$\mment{B}$~(with~$N'<N$
possible results) replaces measurement~$\mment{A}$, the 
probabilities,~$p_1', \dots, p_{N'}'$ of measurement~$\mment{B}$ can
be determined from the~$p_i$ by a many-to-one map of the results
of~$\mment{A}$ to the results of~$\mment{B}$.  For example, in the
case where~$N=3$ and~$N'=2$, the map from the results
of~$\mment{A}$ to the results of~$\mment{B}$ might consist in~$1
\rightarrow 1'$, $2 \rightarrow 2'$ and~$3 \rightarrow 2'$, in which
case~$p_1' = p_1$ and~$p_2' = p_2 + p_3$.   In such a case, we shall
say that measurement~$\mment{B}$ is a \emph{degenerate form} of
measurement~$\mment{A}$; or, more simply, that
measurement~$\mment{B}$ is a degenerate measurement.

Now, measurement~$\mment{B}$ can \emph{formally} be treated as if it
has~$N$ possible results, but where some of these results have the
\emph{same} value.  In this mode of description, in the above
example, one can maintain a one-to-one map between the results
of~$\mment{A}$ and of~$\mment{B}$~(so that~$1 \rightarrow 1'$, $2
\rightarrow 2'$ and so on), but label the results of~$\mment{B}$
with their values, and, when computing the
probabilities of~$\mment{B}$, group together the results which have the
same value.  In the above example, one would
label the three results with values~$b_1, b_2$ and~$b_3$, respectively,
and but have~$b_2=b_3$.

Since measurement~$\mment{B}$ is a degenerate form of
measurement~$\mment{A}$, it can be represented by the
$N$-dimensional degenerate Hermitian operator~$\cmatrix{B} = \sum_i
b_i \cvect{v}_i \cvect{v}_i^\dagger$, where~$\cmatrix{A}\cvect{v}_i
= a_i \cvect{v}_i$.  The probabilities for
measurement~$\mment{B}$ can then be computed in the usual way, on
the understanding that those results with the same values
must not be regarded as physically distinguishable, but must be
grouped as just described.

Conversely, in an abstract set-up where~$\set{A}$ contains
measurements represented by all possible non-degenerate Hermitian
operators, a degenerate Hermitian operator can be regarded as
representing a measurement which is a degenerate form of some
measurement in~$\set{A}$.

\subsection{A Transformation-Invariant Metric and Measure over State Space} \label{sec:Haar}

When the state~$\lvect{Q}$ in Eq.~\eqref{eqn:defn-of-lvect-Q-2} is written in complex form~$\cvect{v} = (Q_1 + iQ_2, \dots, Q_{2N-1} + iQ_{2N})\transpose$, the metric~$ds^2 = \sum_q dQ_q^2$ over~$Q^{2N}$ becomes~$ds^2 = |d\cvect{v}|^2$ in~$\numberfield{C}^N$.   By construction, this metric is invariant under arbitrary unitary or antiunitary transformations.

Furthermore, the metric over~$Q^{2N}$ induces a measure that is uniform over~$\hypersphere$, and that inherits the transformation-invariance of the metric.

\section{Discussion}
\label{sec:discussion}

\subsection{Discussion concerning individual outcomes and their polarities}
\label{sec:discussion-of-postulate-2-2}

The Complementarity and States postulates assert that,
when measurement~$\mment{A}$ is performed, there are~$2N$ possible outcomes, each with an associated
polarity~(a binary degree of freedom), but that, when
outcomes~$ia$ or~$ib$ occur~(with their respective polarities),  only the result~$i$ is registered.  The above derivation of the quantum formalism lends support to the
plausibility of these assertions, but raise the immediate question as
to why the measurement does not~(or cannot) resolve the individual outcomes and their associated polarities, and how one can more intuitively understand their physical meaning. A tentative answer to these questions is as follows.

\subsubsection{Unobservability of the individual outcomes and their polarities.} First, as we shall derive
in Paper~II, the overall phase,~$\phi$, of a system in an eigenstate of energy,~$E$, changes at the rate~$E/\hbar$. Consequently, the probabilities~$P_{2i-1}$ and~$P_{2i}$ of outcomes~$2i-1$ and~$2i$ are oscillating at frequency~$2E/h$, and the polarities of these outcomes are
switching at frequency~$E/h$.  Now, it is reasonable to suppose that,
if one wishes to observe the realization of outcomes~$ia$ or~$ib$ and their polarities, 
the measurement performed must have a
temporal resolution~$\Delta t \ll h/2E$.  Conversely, if the measurement does
not have such resolution, it seems reasonable to suppose that one outcome and its polarity will not 
be cleanly realized, but rather \emph{both} outcomes and their polarities will be realized
over the duration of the measurement, leading to the
situation where only the property that both outcomes hold in common, namely the result~$i$,
can be observed, with the other properties~(namely~$a$ or~$b$, and the polarity~$+$ or~$-$) being `smeared out'.

Now, according to the energy-time uncertainty relation~$\Delta E
\Delta t \geq \hbar/2$~\footnote{%
    We shall regard~$\Delta E \Delta t \geq \hbar/2$ as being a
    consequence of the classical result~$\Delta\omega \Delta t \geq
    1/2$~(relating the uncertainty in the duration and angular
    frequency of a wave) and the photon energy-frequency
    relationship~$E=\hbar \omega$.  However, the validity and meaning
    of the energy-time uncertainty relation, and of the inferences
    that can legitimately drawn from it, have been, and continue to
    be, the subject of debate~(see, for example, Ref.~\cite{Peres-QT}, \S12.8,
    and Ref.~\cite{Oppenheim-PhD}). The
    argument given in the text leading to~$\Delta E \geq E/2$ should,
    accordingly, only be regarded as suggestive insofar as it relies
    on a particular interpretation of the energy-time uncertainty
    relation.},
the energy associated with the interaction used to implement a
measurement with the temporal resolution needed to observe the
individual outcomes and polarities has uncertainty~$\Delta E \geq \frac{1}{2}
\hbar/\Delta t$, so that~$\Delta E \gg E/2\pi$. From~$E=mc^2$, it
then follows that~$\Delta E$ must be of the order of the rest energy
of the system.  A measurement of such energy would therefore
probably not preserve the integrity of the system, thereby violating
the assumption made in Sec.~\ref{sec:F-idealised-set-up} that interactions preserve the integrity of
the system. Hence, a
measurement with the requisite temporal resolution cannot be
consistently described within the quantum formalism. Conversely, a
measurement which, with high probability, preserves the integrity of
the system, will have insufficient temporal resolution to resolve
the individual outcomes and their polarities.

\subsubsection{Physical meaning of the outcomes and polarities}

\begin{figure}[!h]
                \begin{centering}
                \includegraphics[width=2.75in]{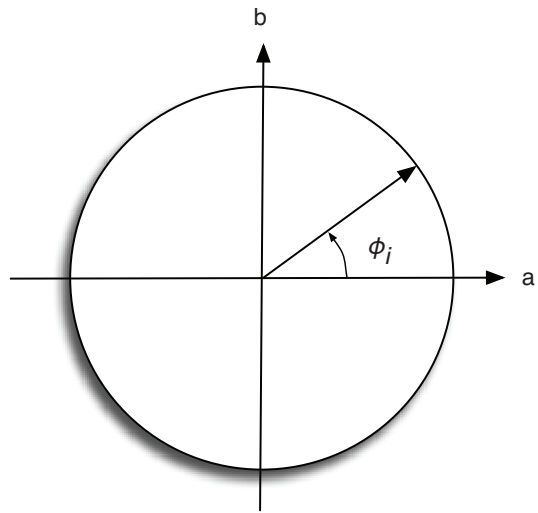}
                \caption{\label{fig:clock} A vector in $ab$-space inclined at angle~$\phi_i$ to the $a$-axis.  When measurement~$\mment{A}$ is performed, outcome~$a$ or~$b$ is obtained with probability~$\cos^2\phi_i $ and~$\sin^2\phi_i$, respectively.}
                \end{centering}
                \end{figure}

The outcomes and polarities can be visualized as follows.  Since the values of the $\phi_i$-dependent state degrees of freedom,~$q_{a|i}$ and~$q_{b|i}$, are unaffected by the addition of a multiple of~$2\pi$ to~$\phi_i$, the physically relevant information about~$\phi_i$ consists of its value mod~$2\pi$.  Furthermore, the measure over~$\phi_i$ is uniform.  Hence, $\phi_i$ can be faithfully represented as shown in Fig.~\ref{fig:clock}.  In the figure, the value of~$\phi_i$ in the range~$[0, 2\pi]$ is shown, and the measure over this interval is uniform, as is implicit in the Euclidean representation.  When measurement~$\mment{A}$ is
performed, the possibilities~$a$ and~$b$ for given~$i$ register which \emph{axis} the vector
is found to be pointing along, and the polarities,~$+$ and~$-$,
determine whether the vector is pointing along the positive
or negative direction along the respective axis.

The probability that~$a$ is obtained is the same as the probability that a photon plane-polarized at angle~$\phi_i$ to the horizontal passes a horizontally-inclined filter, namely~$\cos^2\phi_i$.  This visual representation suggests that the unobserved features of the outcomes~$ia$ and~$ib$ may be connected to Einstein's idea massive particles can be regarded as energy that is `trapped' in a region of space and is undergoing rapid, to-and-fro motion of some kind, which accounts for their inertia~\footnote{See, for example, Ref.~\cite{Bohm-SR}, Ch.~19},  to Hestenes' contention that an electron consists in a localized circular motion which accounts for electron spin~\cite{Hestenes-ZBW1, Hestenes-ZBW2}, and to the Zitterbewegung of an electron that is seen in the solution of the Dirac equation.  The exploration of these tentative suggestions remains for future work.
               
\subsection{Some Implications of the Deduction}

\subsubsection{The Role of Information in Quantum Theory}

One of the major objectives of the program of deriving quantum
theory using the concept of information is to determine whether the
concept of information is indispensable to our understanding of the
quantum formalism, and, if so, to illuminate the precise
relationship between the concept of information and the quantum
formalism.

On the first issue, although many recent approaches to derive the
quantum formalism involve the concept of information, the conclusion
that information is indispensable to our understanding of the
quantum formalism cannot be drawn, either because the approaches are
unable to obtain the quantum formalism~(even though they are able to
derive specific results, such as Malus' law), or because, in those
approaches that are able to obtain a significant fraction of the
quantum formalism,  the abstract nature of some of the assumptions
that are employed obscures the role played by information in
determining the formalism.  Indeed, further doubt on the need for
information is cast by other recent approaches, most notably due to
Hardy~\cite{Hardy01a, Hardy01b}, that are successful in deriving a
significant fraction of the quantum formalism without explicitly
invoking the concept of information.

The formulation presented here provides significant new insight into
this issue.  The formulation rests on assumptions that are
transparent and that are, to a large extent, traceable to familiar
or well-established experimental facts or theoretical idea, so that, for
example, abstract assumptions that directly introduce complex
numbers are avoided.  Consequently, the role played by information
in the derivation can be clearly seen.  Furthermore, one can see
that its role is sufficiently widespread that it seems very likely
that the concept of information could indeed have a fundamental role
to play in our understanding of the origin of the quantum formalism.

On the second issue, owing to the transparency of the assumptions
that form the basis of the derivation, it is apparent from the derivation that the
concept of information, via the information metric, plays a substantial role in giving rise to
various structural features of the quantum formalism.    First, we have seen that, if the space of probability distributions is endowed with the information metric, a transformation to the space where the distributions are parameterized by the square-roots of probability is endowed with a Euclidean metric.   Hence, even at the level of classical probability theory, it appears, just as one sees in the quantum formalism, that the square-roots of probability have a rather fundamental significance

Second, once a quantum state is represented as a unit vector, $\lvect{Q} = (\sqrt{p_1}f(\var_i), \sqrt{p_1}\tilde{f}(\var_i), 	\dots,  \sqrt{p_N}\tilde{f}(\var_N))\transpose$, in a $2N$-dimension real Euclidean space, the imposition that the measure over~$\var_i$ is flat~(which follows from the requirement that the measure induced by the metric over the~$\lvect{Q}$ is consistent with the global gauge condition) leads immediately to the functions~$f(\var_i) = \cos(a\var_i + b)$ and~$\tilde{f}(\var_i) = \sin(a\var_i + b)$.  Hence, the sinusoidal functions into which the phases,~$\phi_i = a\var_i +b$, in a quantum state,~$\cvect{v}=(\sqrt{p_1}e^{i\phi_i}, \dots, \sqrt{p}_N e^{i\phi_N})^\text{\textsf{T}}$, enter can be directly traced to the concept of information.

Third, the requirement that transformations of the state space preserve the metric leads to the conclusion that the transformations must be orthogonal.   As we have shown, it is then only necessary to impose the Gauge Invariance postulate to restrict the set of allowed transformations to a subset of the orthogonal transformations in order to obtain a one-to-one correspondence between the allowed transformations and the set of unitary and antiunitary transformations of a complex vector space.

\subsubsection{Insights into the Quantum Formalism}

The derivation provides a number of significant insights into the
quantum formalism, of which we mention only a few.  
First, we note that, since the development of the quantum formalism,
there has been some uncertainty as to whether the formalism is the
most general formalism for the description of quantum phenomena in
flat space-time. Various possibilities have been suggested for the
generalization of the formalism which, from a purely mathematical
point of view, seem to be plausible, and which may have interesting
physical consequences. For example, the possibility of non-unitary
temporal evolution has been considered by several
authors~\cite{Weinberg89a, Weinberg89b, Herbert82}.  The derivation
given above gives rise to a mathematical structure that is neither
more nor less general than the finite-dimensional abstract quantum
formalism, and thereby lends support to the view that the quantum
formalism is the most general formalism for the description of
quantum phenomena in flat space-time.

Constructively, the formulation has the potential
to allow one to explicitly work out the effect that specific changes
to particular postulates would have upon the quantum formalism.
For example, if one wishes to modify the quantum formalism so as to
allow continuous transformations to be represented by non-unitary
transformations, one could identify which postulates could be plausibly
modified~(in the present example,
the Gauge Invariance Postulate would seem to be the obvious candidate), and
explicitly follow through the changes in the formalism that result.

The second insight is based on the fact that antiunitary transformations are not generally
regarded as an integral part of the abstract quantum formalism~(as
formalized, for instance, by Dirac or von Neumann), but are instead
usually introduced by reference to the theorem of
Wigner mentioned in the Introduction.
Since this theorem rests on particular assumptions, this raises the
question as to whether antiunitary transformations are as
fundamental as unitary transformations. However, since unitary and antiunitary transformations emerge on an 
equal footing in the derivation presented above, the derivation suggests that antiunitary transformations are an equally fundamental part of the quantum formalism.

Third, the prevalence of complex numbers in the quantum formalism is
perhaps its most mysterious mathematical features.  The
emergence of complex numbers in the derivation depends on most of
the postulates, and so is not easy to unravel, but the role of the
global gauge invariance condition is perhaps the most obvious:~prior to the imposition of the two postulates based on this condition~(namely, the Gauge Invariance and the Measure Invariance postulates), the state space is~$\hypersphere$ in a $2N$-dimensional real space, and the set of all possible transformations is the set of orthogonal transformations.  However, these two postulate restrict the set of all possible transformations of~$Q^{2N}$ to a subset of the orthogonal transformations, and thereby allows the set of all possible transformations to be
represented by the set of all unitary and antiunitary transformations of a suitably defined complex vector
space.  Hence, from this perspective, it appears that the use of complex numbers in the quantum formalism is directly tied to the set of possible transformations of state space, and is critically dependent upon the global gauge invariance condition.

Fourth, we have seen that the quantum formalism can be represented in a real form, where a pure state is represented by a unit vector on the unit hypersphere in a real $2N$-dimensional space, and where allowed transformations are a subset of the set of orthogonal transformations of the unit sphere.  Such a simple, easy visualizable representation, which does not appear widely known, may prove useful in certain contexts.  For example, in this representation, one can immediately see that a uniform measure over the unit hypersphere,~$\hypersphere$, is invariant under all orthogonal transformations, and is therefore unitarily-invariant.

Fifth, it has been suggested~\cite{Popescu-Rohrlich97} that the quantum formalism may owe at least a significant part of its structure to the fact that quantum theory permits non-locality and no-signaling to peacefully coexist, and a number of recent reconstructive approaches~\cite{Clifton-Bub-Halvorson03, Hardy01a, Hardy01b} rely upon postulates that concern the behavior of physically separated sub-systems to arrive at the key mathematical features of the quantum formalism.  However, the above derivation suggests that such considerations are not essential to the development of a physical understanding of the core of quantum formalism.

\section{Conclusion}

In this paper, we have shown that the finite-dimensional abstract
quantum formalism~(apart from the explicit form of the temporal
evolution operator) can be derived within the framework of information geometry
using features of quantum phenomena which can be physically understood by reference to experimental observations and general theoretical principles. The derivation illuminates the physical origin of the quantum formalism, suggests that information plays a key role in giving rise to the quantum
formalism, and potentially has significant implications for the
interpretation and proposed modifications of quantum theory.

\begin{acknowledgments}

I am indebted to Steve Gull and Mike Payne for their support, and to Yiton Fu and Tetsuo Amaya for extensive critical comments and suggestions .  I am also indebted to Harvey Brown, Ariel Caticha, Matthew Donald, Chris Fuchs, Suguru Furuta, Lucien Hardy, Lane Hughston, Kevin Knuth, R\"udiger Schach, John Skilling, Lee Smolin, Rob Spekkens, William Wootters, and many others, for discussions and invaluable comments and suggestions.  I would like to thank the Cavendish Laboratory; Trinity College, Cambridge; Wolfson College, Cambridge; and Perimeter Institute for institutional and financial support.    Research at Perimeter Institute is supported in part by the Government of Canada through NSERC and by the Province of Ontario through MEDT.

\end{acknowledgments}

\end{document}